\documentclass[usenatbib]{mn2e}

\usepackage{epsfig}
\usepackage{mathrsfs,aas_macros}

\def\Lya{Ly$\alpha$~}

\def\LCDM{$\Lambda$CDM~}
\def\HI{\hbox{H~$\rm \scriptstyle I\ $}}
\def\HII{\hbox{H~$\rm \scriptstyle II\ $}} 

\def\HeII{\hbox{He~$\rm \scriptstyle II\ $}}

\title[The ionizing emissivity at $z \geq 5$]{The observed
ionization rate of the intergalactic medium and the ionizing emissivity at 
$z \geq 5$: Evidence for a photon starved and extended epoch of reionization}

\author[J.S. Bolton \& M.G. Haehnelt] {James S.
  Bolton$^{1}$\thanks{E-mail:jsb@mpa-garching.mpg.de} \& Martin G.
  Haehnelt$^{2}$ \thanks{E-mail:haehnelt@ast.cam.ac.uk}\\
  $^1$ Max Planck Institut f{\"u}r Astrophysik, Karl-Schwarzschild
  Str. 1, 85748 Garching, Germany \\
  $^2$ Institute of Astronomy, University of Cambridge, Madingley
  Road, Cambridge,  CB3 0HA \\}

\begin{document}

\date{23 July 2007}

\maketitle

\label{firstpage}

\begin{abstract}

Galaxies and quasars are thought to provide the bulk of the photons
responsible for ionizing  the hydrogen in the intergalactic medium
(IGM).  We use a large set of hydrodynamical  simulations, combined
with measurements  of the \Lya opacity of the IGM taken  from the
literature, to obtain robust estimates  of the  photoionization rate
per hydrogen atom at $z=5$ and $6$. We find the photoionization rate
drops by a factor of two and four, respectively, compared to our
recent measurements at $z = 2 - 4$. The number of ionizing photons
emitted by known sources at $z=5$ and $6$, based on an extrapolation
of source numbers below the detection limit and standard assumptions
for the relationship between the ionizing emissivity and  observed
luminosity density at $1500\rm\AA$, are in reasonable agreement with
the photoionization rates inferred from the \Lya forest if the escape
fraction of ionizing photons from galaxies is large ($\ga$ 20 per
cent).  The expected number of ionizing photons from observed sources
at these redshifts therefore appears sufficient to maintain the IGM in
its highly ionized state. Claims to the  contrary may be attributed to
the adoption of an unduly high value for the clumping factor of
ionized hydrogen. Using physically motivated assumptions  for the mean
free path of ionizing photons our measurements of the  photoionization
rate can be turned into an estimate of the ionizing emissivity.  In
comoving units the inferred ionizing emissivity is nearly constant
over the  redshift range $2-6$ and  corresponds to $1.5-3$ photons
emitted per hydrogen atom over a time interval corresponding to the
age of the Universe at $z=6$.  This strongly suggests that the epoch
of reionization  was photon-starved and extended. Completion of
reionization at or before $z=6$ requires either an emissivity which
rises towards higher redshifts or one which remains constant but is
dominated by sources with a rather hard spectral index.  For standard
assumptions, the ionizing emissivity required for completion of
reionization at or before $z=6$ lies at the upper end of recently
reported  values from searches for high redshift galaxies at $z=8-10$.

\end{abstract}
 

\section{Introduction}

Hydrogen in the intergalactic medium (IGM) is maintained in its highly
ionized post-reionization state by the metagalactic ionizing
background.  This radiation is attributed to the integrated UV
emission shortward of the Lyman limit from young star forming galaxies
and quasars, which is subsequently filtered and reprocessed as it
propagates through the clumpy IGM
(\citealt{HaardtMadau96,Fardal98,Madau99}). Observational constraints
on the amplitude and spectral shape of the ionizing background can be
used to infer the relative abundances of different UV photon emitting
populations and their evolution
(\citealt{MiraldaEscudeOstriker90,MeiksinMadau93,Bi97,Devriendt98,Shull99,Bianchi01,Haehnelt01,Shull04,Demianski04,Bolton05,Bolton06}).
The integrated UV emissivity from galaxies can also be directly linked
to the star formation density in the Universe at a particular epoch
(\citealt{Madau98}), and the metagalactic hydrogen ionization rate is
closely related to the evolution of the neutral hydrogen fraction in
the IGM (\citealt{Fan02,Fan06,Becker07}).  Models of the metagalactic
ionizing background are also an essential ingredient within
hydrodynamical simulations of structure formation
(\citealt{Cen92,MiraldaEscude96,Theuns98,Jena05}) and provide the
ionization corrections needed to infer the abundances of heavy
elements in the IGM from metal ion absorption lines
(\citealt{GirouxShull97,Aguirre05,Oppenheimer06}).

Consequently, much effort has been directed towards constraining the
nature and evolution of the ionizing background over a wide range of
redshifts.  Numerical simulations of structure formation, combined
with the \Lya opacity distribution observed in quasar absorption
spectra, have enabled significant progress to be made in this field
(\citealt{Rauch97,McDonaldMiraldaEscude01,MeiksinWhite03,Tytler04,Bolton05,Jena05}).
However, at redshifts approaching and beyond the putative tail-end of
the hydrogen reionization epoch at $z \simeq 6$, determining the
ionization state of the IGM is still problematic ({\it e.g.}
\citealt{Fan02,Santos04,Songaila04,Gallerani05,Fan06,BoltonHaehnelt07,Dijkstra07}).
It is generally agreed that by $z \simeq 6$ the ionizing radiation
emitted by quasars alone is insufficient to reionize the IGM
(\citealt{Fan01b,MiraldaEscude03,Schirber03,Dijkstra04,YanWindhorst04,Meiksin05,Srbinovsky07,Shankar07}).
In addition, there have been some claims that the observed galaxy
population at $z\simeq 6$ is also unable to produce the number of
hydrogen ionizing photons needed to maintain the surrounding IGM in a
highly ionized state (\citealt{LehnertBremer03,Bunker04}).  However,
other studies have indicated this deficit may not real
(\citealt{Stiavelli04,YanWindhorst04b,Bouwens06}), and may instead be
attributable to an overestimate of the number of ionizing photons
required to balance recombinations in the IGM.  There may also be a
population of fainter galaxies which remain undetected.  Considerable
uncertainties are also introduced when converting the galaxy
luminosity function into a UV emissivity and hence an ionizing photon
production rate. An assumption must also be made for the very
uncertain escape fraction of ionizing photons from their respective
galactic environments (\citealt{Giallongo02,FernandezSoto03,Inoue06,Siana07}),
which is likely to vary considerably (\citealt{Shapley06}).

In this paper we shall readdress the question of whether or not the
observed ionization state of the IGM at $z\sim 5-6$ is consistent with
the ionizing emissivity inferred from the observed space density of
star-forming galaxies and quasars.  We approach this issue with
improved measurements of the metagalactic photoionization rate per
hydrogen atom, $\Gamma_{-12}=\Gamma_{\rm HI}/10^{-12}\rm~s^{-1}$, at
$z=5$ and $6$.  These are obtained from recently published
measurements of the \Lya opacity in high redshift quasar spectra
(\citealt{Songaila04,Fan06}), combined with a large suite of high-resolution
hydrodynamical simulations ({\it e.g.}
\citealt{Rauch97,McDonaldMiraldaEscude01,CenMcDonald02,Tytler04,Bolton05}).
These estimates represent the most detailed constraints on
$\Gamma_{-12}$ obtained at these redshifts using state-of-the-art
hydrodynamical simulations, and the systematic uncertainties are
carefully explored for a wide range of model parameters.  These new
data are then compared to the ionization rate inferred from our own
estimates for the ionizing emissivity derived from published
measurements of the space density of Lyman break galaxies (LBGs) and
quasars, combined with a simple model for the mean free path of
ionizing photons.  We also compare this approach to that used by
\cite{Madau99}, who instead compared an estimate of the observed
ionizing emissivity to that required to balance recombinations in a
simple model where the inhomogeneous spatial distribution of the IGM
is characterised by an overall \HII clumping factor. In this way, we
derive an upper limit on the \HII clumping factor from the
$\Gamma_{-12}$ we measure at $z=6$.  Finally, we turn our new
measurements of the ionization rate at $z\sim 5-6$ into a measurement
of the ionizing emissivity.  Given this constraint, we then explore
the range of plausible evolutionary histories for the ionization state
of the IGM towards yet higher redshifts and compare these to recent
observational estimates of the UV emissivity at $z=8-10$.

The structure of this paper is as follows.  In Section 2 we describe
our procedure for estimating the metagalactic hydrogen ionization rate
at $z=5$ and $6$ using measurements of the IGM \Lya opacity and
hydrodynamical simulations of the IGM.  We pay particular
attention to the systematic uncertainties involved in such a
measurement, and as such this work closely follows the methodology
used by \cite{Bolton05} (hereafter B05).  We refer the reader to this
work for further details.  In Section 3 we discuss the model we use
for the mean free path of ionizing photons, and in Section 4 we use
published LBG and quasar luminosity functions to directly compute the
Lyman limit emissivity at $z=5$ and $6$.  We discuss a model for
spatial fluctuations in the ionizing background in Section 5.  This is
used to estimate the possible systematic effect these have on our
measurements of $\Gamma_{-12}$.  In Section 6 we compare the
metagalactic hydrogen ionization rates inferred from both the \Lya
opacity of the IGM and the combined galaxy and quasar Lyman limit
emissivities, and in Section 7 we discuss the corresponding 
comoving ionizing emissivity and its implications for the
ionization state of the IGM at $z>6$. Implications for future 21cm
experiments and the evaporation of mini-haloes are also considered. 
Lastly, in Section 8, we summarise and present our conclusions.


\section{Inferring the ionization rate from the IGM \Lya opacity}
\subsection{Simulations of the IGM}

The hydrodynamical simulations used for this study were run using the
parallel TreeSPH code {\tt GADGET-2} (\citealt{Springel05}).  Our
fiducial simulation volume is a periodic box $15h^{-1}$ comoving Mpc
in length containing $2 \times 200^{3}$ gas and dark matter particles.
We find this provides the best compromise between accuracy and speed.
Star formation is included using a simplified prescription which
converts all gas particles with overdensity $\Delta > 10^{3}$ and
temperature $T<10^{5}\rm~K$ into collisionless stars.   The uniform UV
background model of \cite{HaardtMadau01} (hereafter HM01) including
emission from quasars and galaxies is employed in the optically thin
limit, and the ionization state of the gas is computed using the
non-equilibrium ionization algorithm of \cite{Bolton06}.  The
simulations were all started at $z=99$, with initial conditions
generated using the transfer function of \cite{EisensteinHu99}.  The
cosmological parameters of the simulations used in this study are
listed in Table 1, with the final column listing the factor by which
the \HeII photo-heating rate is artificially raised to explore the
effect of gas temperature on the inferred $\Gamma_{-12}$.  The
cosmological parameters used for the fiducial model are consistent
with the combined analysis of the third year WMAP and \Lya forest data
(\citealt{Viel06,Seljak06}).

To check numerical convergence we also run six further hydrodynamical
simulations with differing box sizes and mass resolutions.  The
different resolution parameters for these simulations are listed
in Table 2.  All other aspects of these simulations are identical to the
fiducial $15-200$ run.

\begin{table}
\centering
\caption{Simulations used in our study of the dependence of
$\Gamma_{-12}$ on various cosmological and astrophysical parameters.
A flat Universe with $\Omega_{\Lambda}=1-\Omega_{\rm m}$ is assumed.
The last column lists the factor by which the \HeII photoheating rate
is multiplied to investigate the effect of gas temperature on the
inferred $\Gamma_{-12}$.  All simulations listed have a box size of
$15h^{-1}$ comoving Mpc and contain $2\times 200^{3}$ gas and dark
matter particles.}
\begin{tabular}
{c|c|c|c|c|c|c|c}
  \hline
    Name & $\Omega_{\rm m}$ & $\Omega_{\rm b} h^{2} $ & $h$ &
  $\sigma_{8}$ & $n$ & $X_{\rm HeII}$ \\  
  \hline
 15-200        & 0.26 & 0.024 & 0.72 & 0.85 & 0.95 & 1   \\
 T1            & 0.26 & 0.024 & 0.72 & 0.85 & 0.95 & 0.2 \\ 
 T2            & 0.26 & 0.024 & 0.72 & 0.85 & 0.95 & 0.5 \\ 
 T3            & 0.26 & 0.024 & 0.72 & 0.85 & 0.95 & 1.5 \\
 T4            & 0.26 & 0.024 & 0.72 & 0.85 & 0.95 & 2   \\
 M1            & 0.17 & 0.024 & 0.72 & 0.85 & 0.95 & 1   \\
 M2            & 0.40 & 0.024 & 0.72 & 0.85 & 0.95 & 1   \\
 M3            & 0.70 & 0.024 & 0.72 & 0.85 & 0.95 & 1   \\
 M4            & 1.00 & 0.024 & 0.72 & 0.85 & 0.95 & 1   \\
 S1            & 0.26 & 0.024 & 0.72 & 0.50 & 0.95 & 1   \\
 S2            & 0.26 & 0.024 & 0.72 & 0.70 & 0.95 & 1   \\
 S3            & 0.26 & 0.024 & 0.72 & 1.00 & 0.95 & 1   \\
 S4            & 0.26 & 0.024 & 0.72 & 1.20 & 0.95 & 1   \\
 \hline
\end{tabular}
\end{table}

\begin{table} 
\centering
  \caption{Resolution and box size of our six additional simulations
  which have the same parameters as the 15-200 model used for
  our parameter study.  The mass resolution used for our parameter
  study simulations is also listed for comparison.}
  \begin{tabular}{c|c|c|c|c|c|c|c}
    \hline
    Name     & Box size       & Total particle         & Gas particle  \\  
             & comoving Mpc   &  number  &  mass $[h^{-1}M_{\odot}]$ \\
  \hline
  15-400     & 15$h^{-1}$     & $2 \times 400^{3}$   & $6.78 \times 10^{5}$ \\
  30-400     & 30$h^{-1}$     & $2 \times 400^{3}$   & $5.42 \times 10^{6}$ \\
  15-100     & 15$h^{-1}$     & $2 \times 100^{3}$   & $4.34 \times 10^{7}$ \\
  30-200     & 30$h^{-1}$     & $2 \times 200^{3}$   & $4.34 \times 10^{7}$ \\
  60-400     & 60$h^{-1}$     & $2 \times 400^{3}$   & $4.34 \times 10^{7}$ \\
  30-100     & 30$h^{-1}$     & $2 \times 100^{3}$   & $3.47 \times 10^{8}$ \\
  \hline
  15-200        & 15$h^{-1}$     & $2 \times 200^{3}$   & $5.42 \times 10^{6}$ \\
  \hline
\end{tabular}
\end{table}

\subsection{Synthetic spectra generation and fiducial parameter ranges}

\begin{figure}
\begin{center} 
  \includegraphics[width=0.45\textwidth]{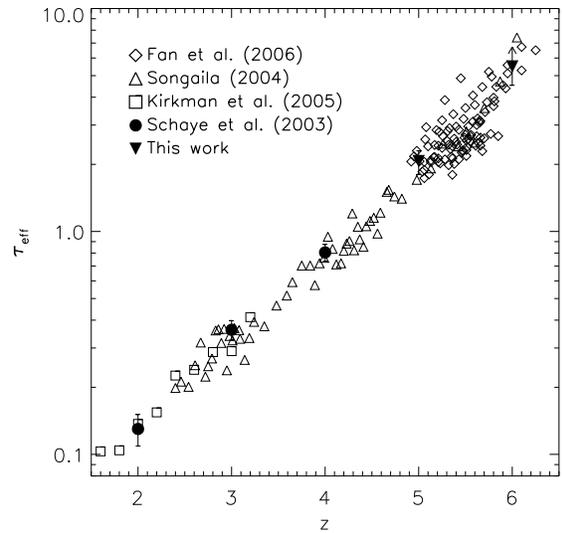}
  \caption{Observational constraints on the \Lya effective optical
  depth of the IGM and its evolution with redshift.  The open squares,
  triangles and diamonds correspond to the data of Kirkman et al. (2005),
  Songaila (2004) and Fan et al. (2006) respectively.  The filled circles
  with error bars show the estimates for $\tau_{\rm eff}$ used in
  B05, based on the data of Schaye et al. (2003).  The filled
  inverted triangles with error bars correspond to the constraints
  on $\tau_{\rm eff}$ we use for this work.}
\label{fig:taueff}
\end{center} 
\end{figure} 

Synthetic \Lya spectra are constructed at $z=5$ and $6$ using $1024$
random lines-of-sight drawn parallel to the box boundaries from each
of the simulations ({\it e.g.}  \citealt{Theuns98}).  Each
line-of-sight consists of $1024$ pixels.   To determine $\Gamma_{-12}$
we rescale the synthetic spectra to match the mean normalised flux,
$\langle F \rangle$, observed in the \Lya forest portion of quasar
spectra at $z=5$ and $6$.  This is achieved by linearly rescaling the
optical depths in each pixel of the synthetic spectra by a constant
factor, $A$, such that

\begin{equation} \langle F \rangle = \frac{1}{N}\sum_{j=1}^{N}
e^{-A\tau_{\rm j}} = e^{-\tau_{\rm eff}}, \end{equation}

\noindent
where $\tau_{\rm j}$ is the optical depth in each of the N pixels in
the synthetic spectra and $\tau_{\rm eff}= -\ln \langle F \rangle$ is
the observed \Lya effective optical depth.  The $\Gamma_{-12}$ required to
reproduce $\tau_{\rm eff}$ is then given by $\Gamma_{-12}=\Gamma_{\rm
sim}/A$, where $\Gamma_{\rm sim}$ is the ionization rate originally
used in the simulation.  Repeating this procedure for many simulations
with varying cosmological and astrophysical parameters allows one to
determine how $\Gamma_{-12}$ scales with these parameters ({\it e.g.}
B05, \citealt{Jena05}).  The values adopted for $\tau_{\rm eff}$ in
this study are based on the recent data published by \cite{Fan06} and
\cite{Songaila04}.  These data are displayed in
Fig.~\ref{fig:taueff} along with the
lower redshift data of \cite{Kirkman05}.  The filled circles
correspond to $\tau_{\rm eff}$ measured by \cite{Schaye03}, used in
B05 for determining $\Gamma_{-12}$ at $2\leq z \leq 4$.

Approaching $z=6$ there is some debate whether an abrupt transition
(\citealt{Fan02,Fan06}) or smooth progression
(\citealt{Songaila04,Becker07}) in the redshift evolution of
$\tau_{\rm eff}$ is a better fit to the observational data.  Rather
than attempt to parameterise the redshift evolution of $\tau_{\rm
eff}$, we bin the data into redshift bins of width $\Delta z = 0.25$
and compute the mean in each bin.  These are shown at $z=5$ and $6$ as
the filled inverted triangles in Fig.~\ref{fig:taueff}, with
corresponding uncertainties estimated from the interquartile range of
the binned data.  Note that at $z\sim6$ some quasar sight-lines
exhibit a full \cite{GunnPeterson65} trough, and therefore only a
lower limit on $\tau_{\rm eff}$, and hence upper limit on
$\Gamma_{-12}$, may be obtained (\citealt{Fan06}).  The fiducial
values for the \Lya effective optical depth adopted for this study are
therefore $\tau_{\rm eff} = 2.07^{+0.23}_{-0.27}$ at $z=5$ and a
lower limit of $\tau_{\rm eff} > 5.50$ at $z=6$, corresponding to
$\langle F \rangle = 0.127^{+0.038}_{-0.027}$ and $\langle F \rangle <
0.004$, respectively.

\begin{table}
\centering
\caption{Fiducial parameter values and estimates of their
uncertainties.  The values listed for $\tau_{\rm eff}$ are at
$z=[5,6]$.}

\begin{tabular}{c|c}
  \hline
    Parameter & Fiducial values and uncertainties\\
    \hline
    $T_{0}$          & $1 \pm0.5 \times 10^{4}$ K   \\
    $\Omega_{\rm m}$         & 0.26$\pm$0.04 \\
    $\tau_{\rm eff}$       & [$2.07^{+0.23}_{-0.27},> 5.50$]  \\
    $\gamma$     & 1.3$\pm$0.3 \\
    $\Omega_{\rm b}h^{2}$         & 0.024$\pm$0.001\\
    $\sigma_{8}$         & 0.85$\pm$0.05  \\
     $h$          & 0.72$\pm$0.04   \\
    \hline
    \label{tab:fiducial}
  \end{tabular}

\end{table}

In addition, we must also make some assumptions for the other
simulation input parameters on which the IGM \Lya opacity depends.
The adopted fiducial values and uncertainties for the cosmological
parameters are $\Omega_{\rm m}=0.26\pm0.04$, $\Omega_{\rm
b}h^{2}=0.024\pm 0.01$, $h=0.72\pm 0.04$ and $\sigma_{8}=0.85\pm
0.05$.  These are consistent with the combined analysis of the third
year WMAP and \Lya forest data (\citealt{Viel06,Seljak06}).
Unfortunately, there are currently no reliable constraints on the
thermal state of the IGM at $z>4$.  However, the slope of the
effective equation of state for the low density IGM,
$T=T_{0}\Delta^{\gamma-1}$,  is likely to be well bracketed by
assuming $\gamma=1.3\pm0.3$ (\citealt{HuiGnedin97,Valageas02}).  We
adopt this as the fiducial range for $\gamma$ in this study.  The
temperature of the IGM at mean density, $T_{0}$, is more difficult to
place limits upon.  We shall assume that the double reionization of
helium does not occur at the same time as hydrogen reionization; there
is evidence for the tail end of \HeII reionization occurring around
$z\simeq 3$ (\citealt{Songaila98,Schaye00,Bernardi03,Bolton06}).
Therefore, $T_{0}$ may reach $1.5 - 2 \times 10^{4}\rm~K$ if hydrogen
and single helium reionization occurred just above $z=6$
(\citealt{HuiHaiman03}), although if hydrogen reionization occurred
earlier $T_{0}$ may be as low as $0.5\times 10^{4} \rm~K$.  We adopt a
fiducial range of  $1 \pm 0.5 \times 10^{4} \rm~K$ for this study.
All our fiducial parameters and their uncertainties are summarised in
Table 3.

\subsection{Numerical resolution}

We must also assess the extent to which changes in the numerical
parameters of the simulations, in this case box size and resolution,
affect the $\Gamma_{-12}$ inferred in our study.  This is achieved by
using the simulations with different box sizes and mass resolutions
listed in Table 2.

Comparing $\Gamma_{-12}$ inferred from the $15-100$ and $60-400$
simulations, which have the same mass resolution,  the larger box size
lowers $\Gamma_{-12}$ by 10 per cent at $z=5$ and 12 per cent at
$z=6$.  A smaller reduction of 4 per cent at both redshifts is found on
comparing the $30-200$ and $60-400$ models.  These results are similar
to those found in B05 over the redshift range $2\leq z \leq 4$.  On
comparing the $15-200$ and $15-400$ simulations which have different
mass resolutions, with increasing resolution $\Gamma_{-12}$ is lowered
by 14 per cent at $z=5$ and 26 percent at $z=6$.  There is
an even larger drop of $26$ per cent and $41$ per cent when comparing
the $15-100$ and $15-200$ models at the same redshifts.  In
comparison, the drop in $\Gamma_{-12}$ found between the $30-200$ and
$30-400$ models is $26$ per cent at $z=5$ and $38$ percent at $z=6$.
Again, the trend in these results is similar to that found in B05, in
the sense that the inferred $\Gamma_{-12}$ decreases with increasing
mass resolution.

Based on these results, we adopt corrections for the box size and mass
resolution which lower the $\Gamma_{-12}$ inferred from the fiducial
$15-200$ simulation by 24 per cent at $z=5$ and 38 per cent at $z=6$.  Note,
however, that the simulations are still only marginally converged.
Therefore, we also adopt an additional numerical uncertainty of 10 per
cent in our final analysis.  From this point onwards all values of
$\Gamma_{-12}$ are quoted including these box size and mass resolution
corrections.

\begin{table} 
\centering
  \caption{The redshift dependent indices for the 
	scaling relations between $\Gamma_{-12}$ and several \
	cosmological and astrophysical parameters.}
  \begin{tabular}{c|c|c|c|c|c}

    \hline
     z    & $T_{0}$ & $\gamma$ & $\Omega_{\rm m}$ & $\sigma_{\rm 8}$ & $\tau_{\rm eff}$ \\
    \hline
     5   & -0.57 & 0.81 & -1.27 & -1.63 & -1.74 \\
     6   & -0.61 & 1.07 & -1.38 & -1.90 & -1.88 \\
  \hline
  
\end{tabular}
\end{table}

\subsection{The dependence of $\Gamma_{-12}$ on cosmological and
astrophysical parameters}

We now determine how $\Gamma_{-12}$ scales with various cosmological
and astrophysical parameters at $z=5$ and $6$ using the simulations
listed in Table 1.  A full discussion of the method used to determine
these scaling relations is outlined in B05. For brevity, in this work
we shall simply report the results of this procedure.  The scaling
relations for $\Gamma_{-12}$ at $z=5$ and $6$ are listed in Table
4. We set $\Gamma_{-12}(z) \propto A^{x(z)}$, where the parameters $A$
are listed in the top row of Table 4 and the relevant indices, $x(z)$,
are listed in the subsequent columns.  Note that, as in B05, we have
adopted $\Gamma_{-12}\propto h^{3}$ and $\Gamma_{-12}\propto
\Omega_{\rm b}^{2}$ independently of redshift.  Additionally, the
scaling of $\Gamma_{-12}$ with the slope of the effective equation of
state, $\gamma$, is derived in post-processing by pivoting the
fiducial $15-200$ model temperature-density relation around the mean gas
density (\citealt{Viel04b}).  This will not self-consistently model
any dynamical effects the change in temperature has on the gas
distribution, but it will model thermal broadening and the change in
the neutral hydrogen fraction correctly, thus providing a reasonable
approximation.  We shall use these derived scaling relations to
estimate the uncertainty on $\Gamma_{-12}$ inferred from our
simulations using the fiducial parameters and uncertainties listed in
Table 3, while also including the corrections for box size and
resolution as discussed above.

\section{From ionization rate to mean free path}

In an inhomogeneous IGM, the mean free path of an ionizing photon is
related to the average separation between Lyman limit systems and the
cumulative opacity of intervening, lower column density \HI absorbers
({\it e.g.}  \citealt{MeiksinMadau93,MiraldaEscude03}).   However, one
may reasonably approximate the ionizing photon mean free path by the
mean separation between Lyman limit systems alone.  Including the
opacity of intervening \HI absorbers will reduce the mean free path
for Lyman limit photons by around a factor of two
(\citealt{MiraldaEscude03,FurlanettoOh05}), but a fraction of photons
at the Lyman limit will also propagate further than one mean free
path.  This will also be true for higher energy photons, reducing the
effect of the intervening opacity further.  Nevertheless, strictly
speaking this assumption should be regarded as providing an upper
limit to the mean free path.

The mean free path model we adopt is based on the post-overlap
reionization model of \cite{MiraldaEscude00} (hereafter MHR00). The
MHR00 model assumes the mean free path for hydrogen ionizing photons,
$\lambda_{\rm mfp}$, corresponds to the mean free path between
regions with overdensity $\Delta>\Delta_{\rm i}$.  The
connection between $\lambda_{\rm mfp}$ and $\Delta_{\rm i}$ is
provided by

\begin{equation} \lambda_{\rm mfp}  = \lambda_{0}(1+z)\left[1-F_{\rm
V}(\Delta<\Delta_{\rm i})\right]^{-2/3},\label{eq:mfp} \end{equation}

\noindent
where $F_{\rm V}(\Delta<\Delta_{\rm i})$ is the fraction of the IGM
volume with $\Delta<\Delta_{\rm i}$, $\lambda_{0}H(z)=60\rm~km~s^{-1}$
and $\lambda_{\rm mfp}$ is expressed as a comoving length.  MHR00 set
$F_{\rm V}(\Delta<\Delta_{\rm i})$ using a fitting formulae for the
volume weighted probability density function (PDF) of the gas
overdensity:

\begin{equation} P_{\rm V}(\Delta)d\Delta = A \exp \left[
-\frac{(\Delta^{-2/3} - C_{0})^{2}}{2(2\delta_{0}/3)^{2}}
 \right]\Delta^{-\beta}d\Delta. \label{eq:densPDF} \end{equation}

\noindent
The fit parameters were derived by MHR00 at $z=2,~3$ and $4$ from the
simulations of \cite{MiraldaEscude96}, and an extrapolation was also
given at $z=6$, assuming $\delta_{0}=7.61/(1+z)$ and a power-law slope
for the high density tail of $\beta = 2.5$.  We have 
checked this extrapolation at $z=6$ against $P_{\rm V}(\Delta)$ derived
from our 15-400 simulation, and find reasonable  agreement at low $\Delta$.
However, even our high-resolution simulations underpredict the amount
of gas at high densities due to limited resolution.  Therefore, in this
work we shall persist in using the $P_{\rm V}(\Delta)$ fits of MHR00,
which include an analytical approximation for the high density tail
of the density PDF as a best guess at correcting for insufficient
numerical resolution.  We also extrapolate $P_{\rm
V}(\Delta)$ to $z=5$ and $z>6$ by assuming $\delta_{0}=7.61/(1+z)$,
$\beta=2.5$ and setting $A$ and $C_{0}$ such that the total volume and
mass is normalised to unity.

In order to link the value of $\Delta_{\rm i}$, and hence
$\lambda_{\rm mfp}$, to a corresponding value for the metagalactic
hydrogen ionization rate, we follow the argument presented by
\cite{FurlanettoOh05}.  Assuming the typical size of an absorber with
overdensity $\Delta$ is the local Jeans length, then in ionization
equilibrium its column density is approximately given by
(\citealt{Schaye01})

\begin{equation} N_{\rm HI} \simeq 4.5 \times 10^{14}
\frac{\Delta^{3/2}}{\Gamma_{-12}} \left( \frac{T}{10^{4}
\rm~K} \right)^{0.2} \left( \frac{1+z}{7} \right)^{9/2} \rm~cm^{-2}. \label{eq:schaye}
\end{equation}

\noindent
The absorber will become optically thick to Lyman limit photons once
$N_{\rm HI} \geq 1/\sigma_{\rm HI} = 1.6 \times 10^{17} \rm~cm^{-2}$,
where $\sigma_{\rm HI}$ is the hydrogen photo-ionization cross-section
at the Lyman limit.  Adopting this value for $N_{\rm HI}$ and
rearranging equation~(\ref{eq:schaye}) yields the expected overdensity
threshold for a Lyman limit system immersed in the metagalactic radiation
field

\begin{equation} \Delta_{\rm i} \simeq 49.5
\left(\frac{T}{10^{4}\rm~K}\right)^{0.13}\left(\frac{1+z}{7}
\right)^{-3} \Gamma_{-12}^{2/3}. \label{eq:deltaSSC} \end{equation}

\noindent
Therefore, given an independent constraint on $\Gamma_{-12}$ we may
estimate $\Delta_{\rm i}$ and hence, using the MHR00 model, the
expected mean free path for hydrogen ionizing photons propagating
through the IGM.  Note, however, that the mean free path in the IGM
can only estimated in this way if it is larger than the mean
separation between ionizing sources ({\it i.e.} once individual
ionized regions have overlapped).  Prior to overlap the sizes of the
ionized regions themselves set the mean free path ({\it e.g.}
\citealt{GnedinFan06}).  Furthermore, before overlap $\Gamma_{-12}$
will vary spatially within each ionized region, and also from one
ionized region to the next, depending on local conditions.  Therefore
the assumption of a monotonic relationship between $\Gamma_{-12}$ and
$\lambda_{\rm mfp}$ will no longer hold.  However, at $z \leq 6$ the
mean free path should still be larger or comparable to the typical
ionizing source separation.

\begin{figure}
\begin{center} 
  \includegraphics[width=0.45\textwidth]{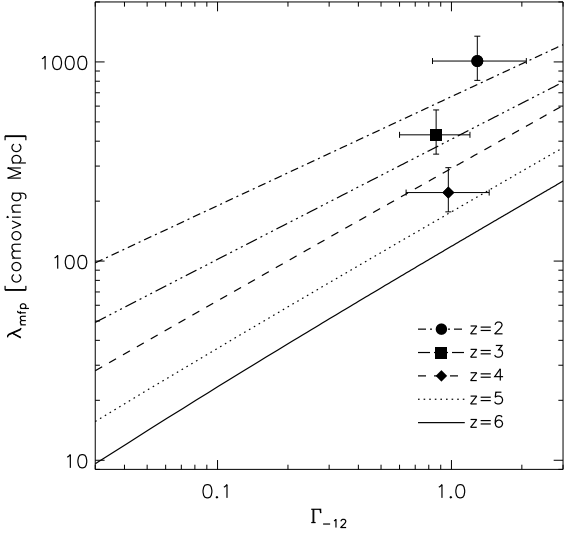}
  \caption{The dependence of the comoving mean free path for ionizing photons
  on the metagalactic hydrogen ionization rate.  The straight lines
  correspond to the predictions made by the model discussed in Section
  3 at several different redshifts.  Independent observational
  constraints on $\lambda_{\rm mfp}$ and $\Gamma_{-12}$ at $z=[2,3,4]$,
  taken from Storrie-Lombardi et al. (1994) and Bolton et al. (2005)
  respectively, are shown by the filled symbols with error bars.}
\label{fig:delta}
\end{center} 
\end{figure} 

The relationship between $\lambda_{\rm mfp}$ and $\Gamma_{-12}$ at
various redshifts is shown as the straight lines in
Fig.~\ref{fig:delta}.  The model is also compared to some independent
observational constraints on $\lambda_{\rm mfp}$ and $\Gamma_{-12}$.
The ionization rate constraints at $z=[2,3,4]$ are taken from B05,
where $\log \Gamma_{\rm HI} =
[-11.89^{+0.21}_{-0.19},-12.07^{+0.17}_{-0.15},-12.01_{-0.18}^{+0.17}]$.
The mean free path estimates are based on the fit to the observed
number of Lyman limit systems per unit redshift at $0.40<z<4.69$ made by
\cite{StorrieLombardi94}.  This yields $dN_{\rm LLS}/dz
=[1.5,2.3,3.3]$ at $z=[2,3,4]$, where

\begin{equation} \lambda_{\rm mfp} \simeq
\frac{c}{H(z)}\left(\frac{dN_{\rm LLS}}{dz}\right)^{-1}. \end{equation}

\noindent
Based on fig. 2 of \cite{StorrieLombardi94}, we assume an uncertainty
of $\pm25$ per cent for the estimate of $dN_{\rm LLS}/dz$ at all
redshifts.  The observational data and the theoretical curves are
consistent with each other within the uncertainties, although the
observed number of Lyman limit systems appears to evolve slightly more
rapidly.


\section{The contribution to the ionizing background from observed galaxies and quasars}

We now proceed to compute the expected contribution to the
ionizing background from LBGs and quasars at
$z=5$ and $6$.  The comoving monochromatic emissivity,
$\epsilon_{\nu}$  $[\rm erg~s^{-1}~Hz^{-1}~Mpc^{-3}]$, of a particular
galaxy or quasar population can be computed from the observed
luminosity function $\phi(L_{\nu},z)$ by

\begin{equation} \epsilon_{\nu} = \int_{L_{\rm
min}}^{\infty} L_{\nu} \phi(L_{\nu},z)dL_{\nu}, \label{eq:epsilon} \end{equation}

\noindent
where in general $L_{\rm min}$ is the luminosity corresponding
to the limiting magnitude of the relevant galaxy or quasar survey;
lower values require the uncertain extrapolation of the faint end of
the luminosity function.  The comoving emissivity is then
related to the metagalactic hydrogen ionization rate by

\begin{equation} \Gamma_{\rm HI} = \lambda_{\rm mfp} (1+z)^{2} f_{\rm esc}
\int_{\nu_{\rm L}}^{\infty} \epsilon_{\nu}  \frac{\sigma_{\nu}}{h_{\rm
p}\nu} d\nu, \label{eq:gammaem} \end{equation}

\noindent
assuming the comoving mean free path, $\lambda_{\rm mfp}$, for all
ionizing photons emitted by a population of Poisson distributed
sources is much smaller than the horizon scale.  This will be a
reasonable approximation at the redshifts we consider (see Table 7 for
details).  Again note that we assume all ionizing photons propagate
for only one mean free path, equivalent to the mean separation between
Lyman limit systems.  Here $f_{\rm esc}$ is the escape fraction of
ionizing photons, $\sigma_{\nu}$ is the photoionization absorption
cross-section, $h_{\rm p}$ is Planck's constant and $\nu_{\rm
L}=13.6\rm~eV/h_{\rm p}$ is the hydrogen Lyman limit frequency.  Under
these assumptions, the comoving emissivity is related to the
specific intensity of the ionizing background, $J_{\nu}$, at frequency
$\nu$ by

\begin{equation} J_{\nu} = \lambda_{\rm mfp}(1+z)^{2} f_{\rm esc} \frac{\epsilon_{\nu}}{4\pi}. \end{equation}

\subsection{Quasar emissivity}

Firstly we consider the expected ionizing emissivity from quasars.  To
estimate the comoving quasar emissivity at $z=5$ and $6$ we have used
the maximum likelihood fits to the quasar luminosity function derived
by \cite{Meiksin05} assuming a pure luminosity evolution,

\begin{equation} \phi(L,z) = \frac{\phi_{*}/L_{*}(z)}{[L/L_{*}(z)]^{-\beta_{1}} + 
  [L/L_{*}(z)]^{-\beta_{2}}}, \end{equation}

\noindent
where  $\beta_{1}=-1.24$, $\beta_{2}=-2.70$, $\rm
log(\phi_{*}/Gpc^{-3}) = 2.51$, $\rm log(L_{*}(5)/L_{\odot})=11.89$
and $\rm log(L_{*}(6)/L_{\odot})=11.53$.  These fits are in good
agreement with the faint and bright end slopes reported by
\cite{Hunt04} at $z\simeq 3$ and \cite{Fan01,Fan04} at $z \geq 5$,
respectively.  However, recent work based on an analysis of deep
X-ray data indicates the faint end slope at $z\sim 6$ may be
significantly steeper (\citealt{Shankar07}).  We adopt a broken power
law for the quasar spectral energy distribution:

\begin{equation}
\epsilon_{\nu} \propto \cases{\nu^{-0.5} &($1050<\lambda<1450\,$\AA),\cr
  \noalign{\vskip3pt}\nu^{-1.5} &($\lambda<1050\,$\AA).\cr}
\end{equation}

\noindent
Assuming the photoionization cross-section has a frequency dependence
$\sigma_{\nu} = 6.3\times 10^{-18} (\nu/\nu_{\rm L})^{-3} \rm~
cm^{2}$, we may then integrate equation~(\ref{eq:gammaem}) to obtain:

\begin{equation} \Gamma_{-12}^{\rm q} \simeq 0.04~ \epsilon_{24}^{\rm q}
  \left( \frac{\alpha_{\rm s}+3}{4.5}\right)^{-1} \left(\frac{\lambda_{\rm mfp}}{40\rm~Mpc}\right)
  \left(\frac{1+z}{7}\right)^{2}, \label{eq:gammaq} \end{equation}

\noindent
where $\epsilon_{24}^{\rm q} = \epsilon^{\rm q}_{\rm L}/10^{24} \rm
~erg~s^{-1}~Hz^{-1}~Mpc^{-3}$ is the comoving quasar emissivity at the
Lyman limit, $\alpha_{\rm s}$ is the spectral index at
$\lambda<912\rm~\AA$ and $\lambda_{\rm mfp}$ is expressed in comoving
Mpc.  We have assumed that all ionizing photons emitted by
quasars escape into the IGM.

There are three main parameters on which $\Gamma_{\rm -12}^{\rm q}$
depends, the ionizing photon mean free path, the spectral index and
the quasar emissivity.  As reiterated  recently  by
\cite{Srbinovsky07}, the emissivity depends sensitively on the minimum
quasar luminosity, $L_{\rm min}$, used when integrating
equation~(\ref{eq:epsilon}).   For this work we shall estimate the
quasar emissivity by integrating equation~(\ref{eq:epsilon}) to a
faint limit of $M_{\rm AB}(1450)=-22$, similar to the faintest
magnitudes probed at these redshifts by combined deep X-ray and
optical surveys ({\it e.g.}
\citealt{Barger03,DijkstraWyithe06,Shankar07}).  This corresponds to
$\log(L_{\rm min}/L_{\odot})=11.17$ within our adopted model.  The
comoving quasar emissivities at the Lyman limit computed using
equation~(\ref{eq:epsilon}) at $z=[5,6]$ are then $\epsilon_{24}^{\rm
q}=[0.57,0.21]$.  These emissivities are similar to those reported by
\cite{Meiksin05}.  These values may be somewhat larger if quasars have
harder spectra at these redshifts. \cite{Scott04} find a harder
spectral index of $\alpha_{\rm s}=-0.56$ may be appropriate below the
Lyman limit, although this value is derived from quasars at $z<1$.

\subsection{Galaxy emissivity}

We compute the expected comoving emissivity from LBGs
in a similar fashion to the quasar emissivity.  LBG luminosity
functions are commonly expressed as a \cite{Schechter76} function, where the
number of galaxies per unit comoving volume is given by

\begin{equation} \phi(L,z) = \frac{\phi_{*}}{L_{*}}
  \left(\frac{L}{L_{*}}\right)^{\alpha}\exp\left(-\frac{L}{L_{*}}\right).
  \end{equation}

\noindent
For the LBG luminosity function at $z=5$, we use the recent data of
\cite{Yoshida06} at $\langle z \rangle =4.7$, where $\alpha=-2.31$,
$\log(\phi_{*}/\rm Gpc^{-3})=5.76$ and $\log(L_{*}/L_{\odot}) =
10.79$.    At $z=6$, we consider two differing, independent
constraints on the LBG luminosity function from \cite{Bouwens06} and
\cite{Bunker04}.  The \cite{Bouwens06} luminosity function parameters
are $\alpha=-1.73$, $\log(\phi_{*}/\rm Gpc^{-3})=6.31$ and
$\log(L_{*}/L_{\odot}) = 10.50$.  The \cite{Bunker04}
data\footnote{Fit taken from table 13 in \cite{Bouwens06}} are
$\alpha=-1.6$, $\log(\phi_{*}/\rm Gpc^{-3})=5.36$ and
$\log(L_{*}/L_{\odot}) = 10.75$.

We adopt a spectral energy distribution based on the model spectrum of
\cite{Leitherer99} for a galaxy of age $500\rm~Myr$ with a continuous
star formation rate,  a Salpeter IMF and metallicity $Z=0.2Z_{\odot}$:

\begin{equation}
\epsilon_{\nu} \propto \cases{\nu^{0} &($912<\lambda<3000\,$\AA),\cr
  \noalign{\vskip3pt}\nu^{-3} &($\lambda<912\,$\AA),\cr}
\end{equation}

\noindent
with an additional break in the spectrum at the Lyman limit,
$\epsilon_{\rm L} = \epsilon(1500)/6$ ({\it e.g.} \citealt{Madau99},
hereafter MHR99).  Once
again, integrating equation~(\ref{eq:gammaem}) we obtain

\[ \Gamma_{\rm -12}^{\rm g} \simeq 0.03~
\epsilon_{25}^{\rm g} \left(\frac{\alpha_{\rm s}+3}{6}\right)^{-1} \left(\frac{f_{\rm esc}}{0.1}
\right) \left( \frac{\lambda_{\rm mfp}}{40\rm~Mpc} \right) \]
\begin{equation} \hspace{6mm} \times \left(
\frac{1+z}{7} \right)^{2}, \label{eq:gammag} \end{equation}

\noindent
where $f_{\rm esc}$ is the escape fraction for ionizing photons,
$\epsilon_{25}^{\rm g} = \epsilon^{\rm g}_{\rm
L}/10^{25}\rm~erg~s^{-1}~Hz^{-1}~Mpc^{-3}$ and $\lambda_{\rm mfp}$ is
again expressed in comoving Mpc.

In this instance there are four parameters which influence estimates
of $\Gamma_{\rm HI}^{\rm g}$: the mean free path, the galaxy
emissivity, the spectral index and additionally the escape
fraction. We attempt to take into account the uncertainty in the
emissivity by integrating equation~(\ref{eq:epsilon}) to two different
lower luminosity limits, roughly corresponding to the faint end limits
of the surveys by \cite{Yoshida06} and \cite{Bouwens06}.  These limits
are $M_{\rm AB}(1350) =-20$ and $-18$ respectively.  Using these
limits, we obtain $\epsilon_{25}^{\rm g} = 0.96$ and $4.31$ for the
\cite{Yoshida06} data at $z=5$.    At $z=6$, we obtain
$\epsilon_{25}^{\rm g} = 0.63$ and $2.36$ for the \cite{Bouwens06}
data and $\epsilon_{25}^{\rm g} = 0.23$ and $0.51$ for the
\cite{Bunker04} data.  These results are summarised along with the
derived quasar emissivities in Table 5.  Lastly, note that the
spectral shape of the ionizing emission from high redshift galaxies is
rather uncertain.  For example, a harder spectrum characterised by
Population-III stars, $\alpha_{\rm s}\sim1$ ({\it e.g.}
\citealt{Bromm01a,Tumlinson03}), would increase the derived ionization
rates by a factor of $1.5$.

\begin{table} 
\centering
  \caption{Summary of the quasar and LBG Lyman limit emissivities derived
  using the luminosity functions and spectral energy distributions
  given in Section 4.  The last column lists the adopted faint end
  magnitude limit, $M_{\rm AB}(1350)$, used when integrating the LBG 
 luminosity function. }
  \begin{tabular}{c|c|c|c|c}
    \hline
    z     & $\epsilon_{24}^{\rm q}$       & $\epsilon_{25}^{\rm g}$  &
  LBG data & magnitude limit \\           
  \hline
 
  5     & 0.57 & 4.31   & Yoshida et al. (2006) & $-18$ \\
        &     & 0.96    & & $-20$ \\
 
  6     & 0.21 & 2.36   & Bouwens et al. (2006) & $-18$\\
        &      & 0.63   & & $-20$ \\
 
  6     & 0.21 & 0.51   & Bunker et al. (2004) & $-18$  \\
        &      & 0.23   & & $-20$  \\

  \hline
\end{tabular}
\end{table}


\section{Spatial fluctuations in the ionizing background approaching reionization}
\subsection{A simple model for ionizing background fluctuations}

\begin{figure*}
  \centering 
  \begin{minipage}{180mm} 
    \begin{center}
         \psfig{figure=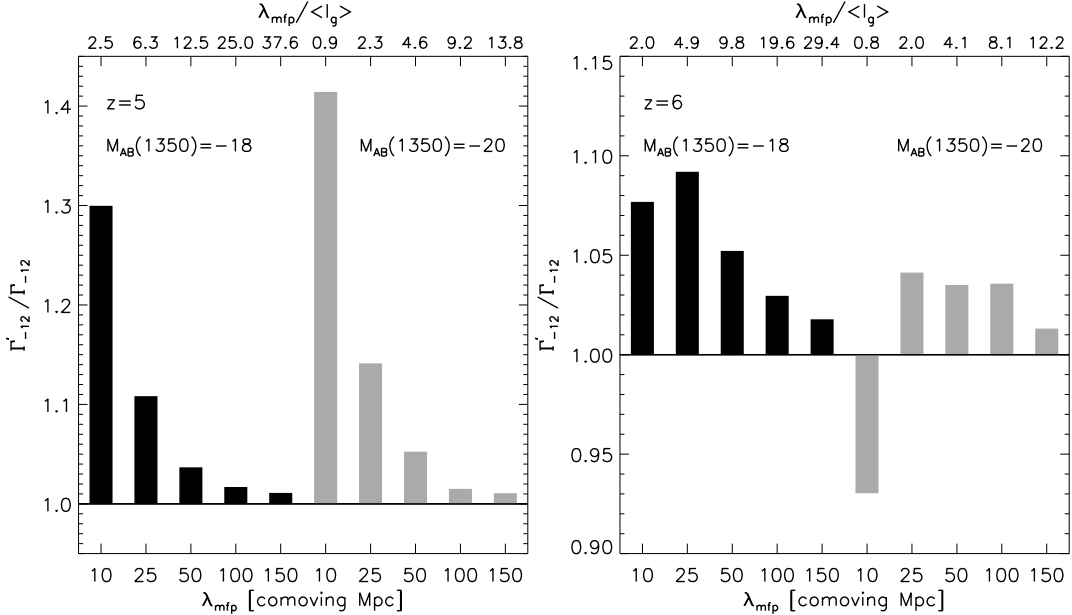,width=0.85\textwidth}
      \caption{{\it Left:} The ratio of the hydrogen ionization rate
      inferred from hydrodynamical simulations with a spatially
      fluctuating ionizing background to that obtained assuming a
      spatially uniform ionizing background at $z=5$.  The lower
      horizontal axis denotes the mean free path used in each
      fluctuating background model, while the upper horizontal axis
      corresponds to the ratio of the mean free path to the mean
      separation between galaxies in each model.  The data are
      constructed from galaxy luminosity functions integrated to
      two different limiting magnitudes, $M_{\rm AB}(1350)=-18$
  and $-20$, shown by the black and grey bars respectively.  {\it Right:} As for left panel but at $z=6$.}
\label{fig:fluctuations} 
\end{center} 
\end{minipage}
\end{figure*}

At $z<4$ spatial fluctuations in the metagalactic hydrogen ionization
rate are expected to be small
(\citealt{Croft99,MeiksinWhite04,Croft04}; but see also
\citealt{Maselli05}).  At these redshifts, the mean free path for
ionizing photons is substantially larger than the mean separation
between ionizing sources, therefore a spatially uniform ionizing
background is expected to be a reasonable approximation.  However,
towards higher redshifts, and especially approaching the tail end of
the hydrogen reionization epoch, spatial fluctuations in the ionizing
background are gradually amplified.  This amplification is
attributable to diminishing sources numbers, a smaller mean free path
and the inhomogeneous distribution of the ionizing sources themselves
({\it e.g.}  \citealt{Bolton06,Wyithe06}).

However, when obtaining $\Gamma_{-12}$ from our simulations we assume
a spatially uniform ionizing background.  We must therefore account
for the systematic error this assumption will impart to our
measurements of $\Gamma_{-12}$.  A correct numerical treatment of
ionizing background fluctuations requires a prescription for
cosmological radiative transfer, which is beyond the scope of this
paper. We instead use a modified version of the fluctuating ionizing
background model of \cite{Bolton06}, originally used to analyse
spatial fluctuations in the quasar dominated \HeII ionizing background
at $2<z<3$, combined with our 30-400 simulation.  The details of the
model implementation are given in \cite{Bolton06}, to which we refer
the reader.  Here we only note the modifications made to the model for
this work.

Firstly, we now use the model to compute fluctuations in the ionizing
background for \HI rather than \HeII ionizing photons.  Therefore,
galaxies as well as quasars are included in the model by using the
LBG luminosity function of \cite{Yoshida06} at $z=5$ and
\cite{Bouwens06} at $z=6$. The hydrogen ionization rate at some
position ${\bf r}_{0}$ may be written as

\begin{equation} \Gamma_{\rm HI} =  \sum_{j=1}^{N} 
\left[ \int^{\infty}_{\nu_{L}} \frac{L^{\rm j}_{\nu} f_{\rm esc}}{4\pi |{\bf
r}_{j}-{\bf r}_{0}|^{2} } \frac{\sigma_{\nu}}{h_{\rm p} \nu}~d\nu \right],  
\label{eq:flucs} \end{equation}

\noindent
for $|{\bf r}_{\rm j}-{\bf r}_{0}| < \lambda_{\rm mfp}/(1+z)$, where
 $|{\bf r}_{\rm j}-{\bf r}_{0}|$ is the proper distance of ionizing
 source $\rm j$ with luminosity $L^{j}_{\nu}$ from ${\bf r}_{0}$ and
 $N$ is the total number of ionizing sources brighter than $L_{\rm
 lim}$ in some fixed volume.  We set $f_{\rm esc}=1$ for quasars,
 $f_{\rm esc}=0.1$ for galaxies and assume that all ionizing photons
 propagate a distance corresponding to one mean free path.  The total
 number of ionizing sources brighter than absolute magnitude $M_{\rm
 lim}$, corresponding to $L_{\rm lim}$, within a volume V at redshift
 $z$ is then fixed by

\begin{equation} N(z,M< M_{\rm lim}) = V \int^{M_{\rm lim}}_{-\infty}
\phi(M,z)~dM, \label{Mlim2N} \end{equation}

\noindent
where $\phi(M,z)$ is the relevant quasar or LBG luminosity function
expressed in terms of magnitudes.  The LBG luminosity functions are
integrated to two different limiting magnitudes, $M_{\rm
AB}(1350)=-18$ and $-20$.  The brighter limiting magnitude produces
fewer galaxies within a fixed volume, increasing the amplitude of the
spatial fluctuations.  The limiting magnitude for the quasar
luminosity function is taken to be $M_{\rm AB}(1450)=-22$ as before.

Haloes are identified within the 30-400 simulation volume using a
friends of friends halo finding algorithm with a linking length of
$0.2$.  Ionizing source luminosities are assigned to the identified
haloes by Monte-Carlo sampling the quasar and LBG luminosity
functions.  The haloes are populated by associating the brightest
sources with the most massive haloes in the simulation volume.  A more
detailed approach is impossible in this instance due to the limited
resolution of the simulation, but this approach should model the
spatial distribution of the ionizing sources reasonably well.  The
galaxy and quasar spectral energy distributions adopted are given
Section 4 of this paper.  The fluctuating ionizing background models
are then constructed at $z=5$ and $6$ using five different values for
the mean free path, $\lambda_{\rm mfp}=10$, $25$, $50$, $100$ and
$150$ comoving Mpc.  The effect of finite source lifetimes and light
cone effects ({\it e.g.}  \citealt{Croft04}) are not included in our
model, but their impact on the fluctuation amplitude for the small
mean free paths considered here is expected to be small
(\citealt{Wyithe06}).

\subsection{The impact of ionizing background fluctuations on
  $\Gamma_{-12}$ inferred from simulations}

The metagalactic ionization rate we infer from our simulations
using a spatially fluctuating ionizing background,
$\Gamma_{-12}^{\prime}$, is compared to $\Gamma_{-12}$ obtained
assuming a spatially uniform ionizing background in
Fig.~\ref{fig:fluctuations}.  The left panel shows the $z=5$ data,
with $\Gamma_{\rm -12}^{\prime}/\Gamma_{\rm -12}$ for each fluctuating
ionizing background model plotted as the black bars for a limiting
magnitude of $M_{\rm AB}(1350)=-18$ and grey bars for $M_{\rm
AB}(1350)=-20$.  The lower horizontal axis corresponds to the
comoving ionizing photon mean free path, while the upper horizontal
axis shows the ratio of the mean free path to the mean separation
between galaxies in the model, $\langle l_{\rm g} \rangle$.  Similar
data at $z=6$ are shown in the right hand panel of
Fig.~\ref{fig:fluctuations}.  Note that $\langle l_{\rm g}
\rangle$ is an average galaxy separation computed
assuming the galaxies are uniformly distributed.  In reality the
inhomogeneous distribution of ionizing sources within our model
results in underdense regions being surrounded by fewer ionizing
sources than overdense regions, further increasing the amplitude of
the spatial fluctuations.

As one might expect, the impact of spatial fluctuations is largest
when $\lambda_{\rm mfp}/\langle l_{\rm g} \rangle$ is small.
Increasing the number of ionizing sources within one mean free path of
a fixed point (either by increasing $\lambda_{\rm mfp}$ or adopting a
fainter limiting magnitude) reduces the amplitude of the fluctuations
and thus the departure of $\Gamma_{-12}^{\prime}$ from the ionization
rate inferred assuming a spatially uniform ionizing background.  As
found in previous studies
(\citealt{GnedinHamilton02,MeiksinWhite04,Bolton06}) the assumption of
a spatially uniform ionizing background tends to underestimate the
ionization rate required to reproduce the observed \Lya effective
optical depth at high redshifts.  The presence of large ionization
rates in regions where the \Lya opacity of the IGM is already rather
low or very high increases the mean $\Gamma_{-12}$ but has little impact on the
observed effective optical depth.  For plausible values for the mean
free path at $z=5$ this effect is around the $5-10$ per cent level.

Interestingly,  there is a different trend at $z=6$ for progressively
smaller values of $\lambda_{\rm mfp}$.  The difference between
$\Gamma_{-12}^{\prime}$ and $\Gamma_{-12}$ is reduced as $\lambda_{\rm
mfp}$ decreases, and $\Gamma_{-12}$ is actually larger than
$\Gamma_{-12}^{\prime}$ for the model with the smallest value of
$\lambda_{\rm mfp}/\langle l_{\rm g} \rangle$ at $z=6$.  This
result was also noted by \cite{MeiksinWhite04}.  In this instance,
the volume of the IGM where there are no ionizing sources within a
mean free path become more common, reducing the
spatially averaged ionization rate.  

In summary, we find the ionization rate inferred from simulations of
the IGM which assume a uniform ionizing background at $z=5$ and $6$ is
usually underestimated.  However, as regions which are isolated from
ionizing photons become more common, the $\Gamma_{-12}$ can actually
be overestimated when assuming a uniform ionizing background.  For
plausible values of $\lambda_{\rm mfp}$, spatial fluctuations in the
ionizing background induce a systematic offset of about $+10$ per cent
and an uncertainty of about $\pm 10$ per cent to the $\Gamma_{-12}$
we infer from our simulations at $z=5$ and $6$, respectively.   We
shall include these uncertainties when quoting our final estimate of
$\Gamma_{-12}$ in Section~\ref{sec:results}.

\section{The metagalactic hydrogen ionization rate at z=5 and 6} \label{sec:results}
\subsection{Constraints on $\Gamma_{-12}$ from the IGM \Lya opacity}

\begin{figure}
\begin{center}
 
  \includegraphics[width=0.45\textwidth]{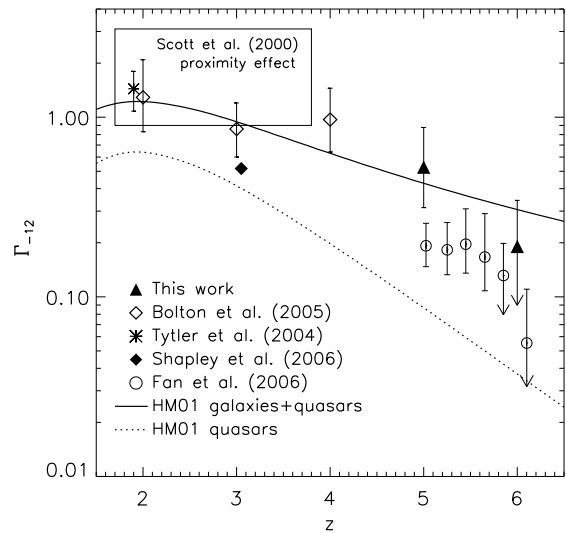}
  \caption{Evolution of the metagalactic hydrogen ionization rate over
  the redshift range $ 2 \leq z \leq 6$.  Our new estimates of 
  $\Gamma_{-12}$ are shown as the filled triangles at $z=5$ and $6$.
  Other data points based on the IGM \Lya effective optical depth are
  taken from Tytler et al. (2004) (star), Bolton et al. (2005) (open
  diamonds) and Fan et al. (2006) (open circles).  The box corresponds
  to the constraints on $\Gamma_{-12}$ obtained from the proximity
  effect by Scott et al. (2000), and the filled diamond is the recent
  estimate of $\Gamma_{-12}$ based on emission from Lyman break
  galaxies at $z=3$ (Shapley et al. 2006).  The solid and dotted lines
  correspond to the ionizing background models of HM01 for,
  respectively, galaxies and quasars and quasars only.}
\label{fig:update}
\end{center} 
\end{figure}

We summarise our constraints on the metagalactic hydrogen ionization
rate at $z=5$ and $6$ in Fig.~\ref{fig:update}.  The values of
$\Gamma_{-12}$ which reproduce the IGM \Lya effective optical depth
are plotted as the filled triangles with $1\sigma$ error bars.  Our central
estimates and their uncertainties are based on $\Gamma_{-12}$ inferred
from our fiducial simulation (15-200), rescaled to correspond to the
fiducial parameter values and ranges listed in Table 3.  We do this
using the scaling relations we derived for $\Gamma_{-12}$ in Section
2.  The total uncertainty on $\Gamma_{-12}$ is then computed by adding
the contributions from individual parameter uncertainties in
quadrature.  The total error budget is summarised in Table 6.  We find
the \Lya effective optical depth is reproduced by $\log \Gamma_{\rm
HI} = -12.28^{+0.22}_{- 0.23}$ at $z=5$ and $\log \Gamma_{\rm HI}<
-12.72$ at $z=6$.

\begin{table} 
\centering
\caption{The error budget for $\Gamma_{-12}$ based on the
fiducial values for various cosmological and astrophysical parameters
given in Table 3.  The uncertainties due to the impact of ionizing
background fluctuations and marginal numerical convergence are also
listed.  The total error is obtained by adding the individual errors
in quadrature. }

\begin{tabular}{c|c|c}
  \hline
    Parameter & $z=5$ (per cent) & $z=6$ (per cent)\\  
  \hline
    $T_{0}$   & ${+49}/{-21}$  & ${+53}/{-22}$   \\ 
    $\tau_{\rm eff}$ & ${+27}/{-17}$  & ${+44}/{-32}$  \\
    $\Omega_{\rm m}$ & ${+24}/{-17}$ & ${+26}/{-18}$    \\
    $\gamma$  & ${+18}/{-19}$  & ${+24}/{-19}$   \\
    $\sigma_{8}$ & ${+10}/{-9}$  & ${+12}/{-10}$  \\
    Fluctuations &  ${+10}$ &  $\pm 10$   \\ 
    Numerical &  $\pm 10$ &  $\pm 10$   \\
    $\Omega_{\rm b}h^{2}$ & ${+9}/{-8}$  & ${+9}/{-8}$    \\
    $h$ &  $\pm 6$ &  $\pm 6$  \\ 
    Total & ${+66}/{-40}$ & ${+80}/{-53}$  \\
    \hline  
    \label{tab:errors}
  \end{tabular}

\end{table}

These data are compared to other observational constraints taken from
the literature in Fig.~\ref{fig:update}.  The open circles with error
bars correspond to the data of \cite{Fan06}, derived from the IGM \Lya
effective optical depth at $z\simeq 5-6$.  Rather than using
hydrodynamical simulations, \cite{Fan06} compute $\Gamma_{-12}$ using
the fluctuating Gunn-Peterson approximation (FGPA,
\citealt{Rauch97,Weinberg99}) combined with the IGM density
distribution derived by MHR00.  The data are in agreement with our
constraints at $z=6$, although our upper limit on $\Gamma_{-12}$ is
somewhat higher.  However, at $z=5$ our determination of
$\Gamma_{-12}$ is significantly larger than the data of \cite{Fan06}.
This is not unexpected; the FGPA should be used with caution, as in
practice the dependence of the \Lya optical depth (not to be confused
with $\tau_{\rm eff}$) on various parameters is actually rather
different to the canonical FGPA scaling (B05,
\citealt{Tytler04,Jena05}).  We therefore expect our constraints on
$\Gamma_{-12}$, based on our extensive suite of hydrodynamical
simulations, to be more robust.  The curves in Fig.~\ref{fig:update}
correspond to the $\Gamma_{-12}$ predicted by the ionizing background
models of HM01 for galaxies and quasars (solid line) and quasars only
(dotted line).  These models are described fully in B05.  Our data are
in good agreement with the model based on the combined emission from
galaxies and quasars, which is consistent with the interpretation that
quasars alone cannot maintain the IGM in its highly ionized state at
$z=6$ (\citealt{MiraldaEscude03,Meiksin05,Srbinovsky07}).  We also
compare these curves to lower redshift constraints on $\Gamma_{-12}$
based on the \Lya effective optical depth.  The open diamonds are the
data from B05 and the star is taken from \cite{Tytler04}.
Measurements of $\Gamma_{-12}$ can also be made using the quasar
proximity effect ({\it e.g.}
\citealt{Bajtlik88,Giallongo96,Scott00}), although these estimates can
be subject to a bias introduced by the overdense regions in which
quasars reside (\citealt{Rollinde05,Guimaraes07,Faucher07}).  The box
in Fig.~\ref{fig:update} corresponds to the $\Gamma_{-12}$ inferred by
\cite{Scott00}.  All data are consistent with the ionizing background
having a substantial contribution from young star forming galaxies.
The filled diamond also shows the recent direct estimate by
\cite{Shapley06} of the contribution of LBGs to the ionizing
background at $z \simeq 3$.  The measurements are consistent with
around $50$ per cent of the ionizing photons at $z\simeq 3$
originating from galaxies.

Lastly, we compare our data with previous constraints on
$\Gamma_{-12}$ obtained from other simulations of the \Lya forest.
\cite{McDonaldMiraldaEscude01} use the hydro-PM code of
\cite{GnedinHui98} to run a simulation with a box size of $8.9h^{-1}$
comoving Mpc and a $256^{3}$ grid.  They obtain $\Gamma_{-12}=0.13\pm
0.03$ at $z=4.93$ assuming a flat \LCDM cosmological model with
$\Omega_{\rm m}=0.4$, $\Omega_{\rm b}h^{2}=0.02$, $h=0.65$,
$\sigma_{8}=0.79$ and $n=0.95$ for $\tau_{\rm eff}=2.3$, $T_{0}=2
\times 10^{4}\rm~K$ and $\gamma=1$.  We rescale their constraint on
$\Gamma_{-12}$ to match our fiducial cosmological parameters, listed
in Table 3, with the scaling relations derived at $z=5$ in Table 4.
\cite{McDonaldMiraldaEscude01} do not apply a numerical resolution
correction factor to their data, so for a fair comparison with our
data we also multiply their rescaled $\Gamma_{-12}$ constraint by the
appropriate resolution correction factor of $0.76$ we obtained at
$z=5$. This procedure yields $\Gamma_{-12} = 0.44\pm0.10$.
\cite{MeiksinWhite04} employ a similar pseudo-hydrodynamical approach
with a simulation box size of $25h^{-1}$ comoving Mpc, $512^{3}$ dark
matter particles and a $1024^{3}$ force mesh.  They obtain
$\Gamma_{-12}=0.31^{+0.07}_{-0.09}$ at $z=5$ for a flat \LCDM model
with $\Omega_{\rm m}=0.3$, $\Omega_{\rm b}h^{2}=0.022$, $h=0.70$,
$\sigma_{8}=0.92$, $n=0.95$ and $\tau_{\rm eff}=2.12$.  The gas is
assumed to follow an effective equation of state with $T_{0}=2\times
10^{4}\rm~K$ and $\gamma=1.5$.  Rescaling their result yields
$\Gamma_{-12}=0.52^{+0.11}_{-0.15}$.  Both of these results are good
agreement with our constraint of $\Gamma_{-12} =
0.52^{+0.35}_{-0.21}$.  At $z=6$, \cite{CenMcDonald02} use an Eulerian
hydrodynamical simulation with a box size of $25h^{-1}$ comoving Mpc
and a $768^{3}$ grid to obtain a $1\sigma$ upper limit of
$\Gamma_{-12}<0.096$.  They adopt $\Omega_{\rm m}=0.3$, $\Omega_{\rm
b}h^{2}=0.02$,  $h=0.67$, $\sigma_{8}=0.9$ and $n=1$ with $\tau_{\rm
eff}=5.57$; no values for $T_{0}$ or $\gamma$ were quoted.  Rescaling
their constraint on $\Gamma_{-12}$ as before, and including an
appropriate resolution correction factor of $0.62$ at $z=6$ yields
$\Gamma_{-12}<0.11$ assuming $T_{0}=10^{4}\rm~K$ and $\gamma=1.3$.
\cite{MeiksinWhite04} quote a $1\sigma$ upper limit of
$\Gamma_{-12}<0.14$ for $\tau_{\rm eff}>5.12$ at $z=6$ .  Rescaling
their result gives $\Gamma_{-12}<0.20$.  These results are again
consistent with our upper limit of $\Gamma_{-12}<0.19$.

\subsection{Constraints on $\Gamma_{-12}$ from the observed galaxy and quasar populations}

\begin{figure*}
  \centering 
  \begin{minipage}{180mm} 
    \begin{center}
           
      \psfig{figure=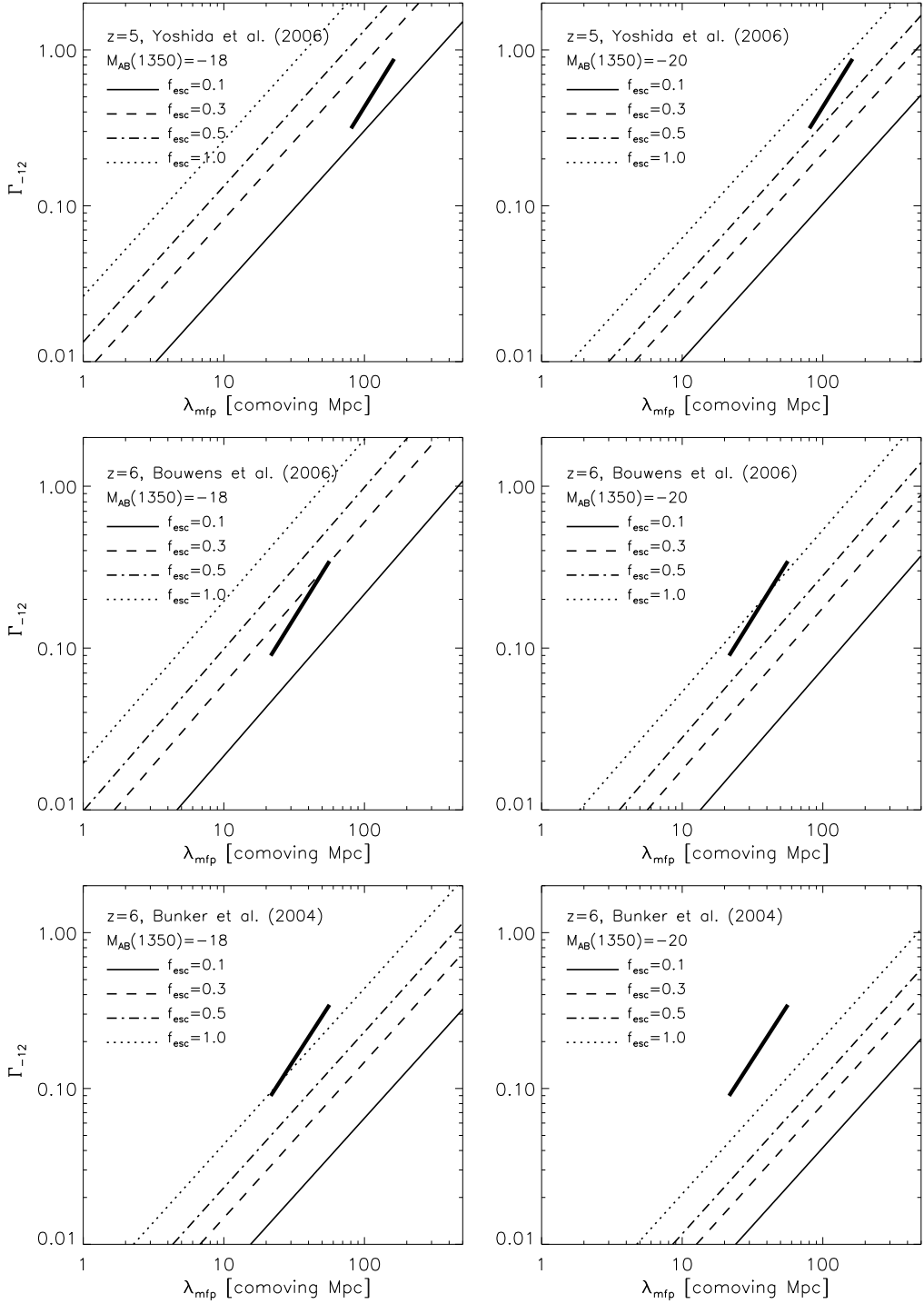,width=0.8\textwidth}
      \caption{Comparison of our constraints on $\Gamma_{-12}$ and
      $\lambda_{\rm mfp}$ obtained from the IGM \Lya effective optical
      depth at $z=5$
      and $6$ (thick solid lines) to $\Gamma_{-12}$ computed from the
      Lyman limit emissivity of the observed galaxy and quasar
      populations at the same redshifts (thin straight lines).  These
      are plotted as a function of $\lambda_{\rm mfp}$ for four
      different values of the ionizing photon escape fraction  from LBGs.
      {\it Top panels:} Data at $z=5$ computed using the LBG
      luminosity function of Yoshida et al. (2006).  The left and
      right hand panel correspond to the LBG emissivities computed
      by integrating the luminosity function to limiting
      magnitudes of $M_{\rm AB}(1350)=-18$ and $-20$
      respectively. {\it Central panels:} Data at $z=6$ computed from
      the LBG luminosity function of Bouwens et al. (2006). The
      left and right hand panels correspond to the galaxy emissivities
      computed by integrating the luminosity function to limiting
      magnitudes of $M_{\rm AB}(1350)=-18$ and $-20$
      respectively. {\it Bottom panels:} As for central panels but
      using the galaxy luminosity function measured  by Bunker et
      al. (2004).  } \label{fig:emissivity} \end{center} \end{minipage}
\end{figure*}

We now compare our estimates for $\Gamma_{-12}$ at $z=5$ and $6$ to
the ionization rates estimated from the Lyman limit emissivities
listed in Table 5.  The total ionization rate is $\Gamma_{-12} =
\Gamma_{-12}^{\rm q} + \Gamma_{-12}^{\rm g}$, where the quasar and
galaxy contributions are computed using equations~(\ref{eq:gammaq})
and (\ref{eq:gammag}).  This is plotted as a function of $\lambda_{\rm
mfp}$ in Fig.~\ref{fig:emissivity} for four different values of the
LBG escape fraction, $f_{\rm esc}$.  The escape fraction for ionizing
photons is very uncertain, although recent work suggests $f_{\rm esc}$
may vary with redshift (\citealt{Razoumov06}), and that $f_{\rm esc} >
0.1$ may be appropriate at $z\simeq 6$ (\citealt{Inoue06})

The top left and right panels in Fig.~\ref{fig:emissivity} display
$\Gamma_{-12}$ obtained by integrating the \cite{Yoshida06} LBG
luminosity function to $M_{\rm AB}(1350)=-18$ and $M_{\rm
AB}(1350)=-20$, respectively.  The limiting magnitude of the
\cite{Yoshida06} data is roughly $M_{\rm AB}(1350)=-20$.  The quasar
contributions to $\Gamma_{-12}$ correspond to 15 per cent for $M_{\rm
AB}(1350)=-18$ and 44 per cent for $M_{\rm AB}(1350)=-20$, assuming
$f_{\rm esc}=0.1$.  For comparison, the thick solid lines show our
determination of $\Gamma_{-12}$ from the IGM \Lya effective optical
depth at $z=5$, with our corresponding estimate for $\lambda_{\rm
mfp}$ using the model described in Section 3.  For $\Gamma_{-12}=
0.52^{+0.35}_{-0.21}$ we obtain a mean free path of $\lambda_{\rm
mfp}=114^{+49}_{-34}$ comoving Mpc. The combined emission from
galaxies and quasars at $z=5$ is consistent with the IGM \Lya
effective optical depth, assuming $M_{\rm AB}(1350)=-18$ and $f_{\rm
esc} \geq 0.15$, or $M_{\rm AB}(1350)=-20$ and $f_{\rm esc} \geq
0.6$.

The central panels show similar data for $\Gamma_{-12}$ at $z=6$ using
the LBG luminosity function of \cite{Bouwens06}.   A constant quasar
contribution is again added, corresponding to 11 percent of the total
$\Gamma_{-12}$ assuming a magnitude limit of $M_{\rm AB}(1350)=-18$
and 31 per cent for $M_{\rm AB}(1350)=-20$, both for $f_{\rm
esc}=0.1$.  The limiting magnitude of the \cite{Bouwens06} data
roughly corresponds to $M_{\rm AB}(1350)=-18$.  The thick solid line
gives our upper limit of $\Gamma_{-12}<0.19$ and
$\lambda_{\rm mfp}<37$ comoving Mpc.  Galaxies and quasars provide the
required number of photons needed to reproduce the IGM \Lya effective
optical depth for $M_{\rm AB}(1350)=-18$ and $f_{\rm esc}\geq 0.2$.
For the \cite{Bouwens06} data there does not appear  to be any deficit
in the number of ionizing photons required to ionize the IGM at $z=6$.
A similar conclusion was also reached by \cite{Bouwens06}.

The bottom panels show $\Gamma_{-12}$ obtained from the LBG luminosity
function of \cite{Bunker04} at $z=6$.  A constant quasar contribution
of 36 percent of the total $\Gamma_{-12}$ is added for $M_{\rm
AB}(1350)=-18$ and 55 per cent for $M_{\rm AB}(1350)=-20$, assuming
$f_{\rm esc}=0.1$.  The \cite{Bunker04} luminosity function has a
substantially lower galaxy number density at the faint end compared to
the \cite{Bouwens06} data.  Consequently, the $\Gamma_{-12}$ computed
from the LBG and quasar emissivities require a much higher value of
$f_{\rm esc}$ for consistency with our constraints on $\Gamma_{-12}$
and $\lambda_{\rm mfp}$ from the IGM \Lya effective optical depth.  At
$M_{\rm AB}(1350)=-18$, roughly corresponding to the limiting
magnitude of the \cite{Bunker04} galaxy luminosity function, $f_{\rm
esc}\sim 1$ is required, whereas for $M_{\rm AB}(1350)=-20$ LBGs and
quasars cannot account for the required ionization rate at $z=6$.  An
escape fraction of $f_{\rm esc}\sim 1$ is probably implausibly high
but given the rather uncertain spectral energy distribution of the
sources this is probably not a cause for concern. 

However, this time our conclusion is rather different from the one
reached by \cite{Bunker04}.  On the basis of their luminosity function
they concluded that there is a significant deficit in the number
ionizing photons at $z=6$ required to keep the IGM highly ionized.
Even for $f_{\rm esc}=1$, Bunker et al.  found the number of photons
required  to ionize the IGM provided by galaxies suffers a substantial
shortfall. Note that this difference is not largely due to our
assumption that quasars also contribute to the ionizing emissivity.
For the assumed emissivities given in Table 5, quasars only increase
the total number of ionizing photons by around a factor of two
assuming an LBG escape fraction of $f_{\rm esc}=0.1$, and by much
smaller factors for larger escape fractions.  As discussed in the next
section, this discrepancy can be attributed to the assumption of an
unduly large  \HII  clumping factor when estimating the number of
photons required to keep the IGM highly ionized.

Lastly, one may have some concern over the above conclusions regarding
the adopted mean free path. Recent cosmological radiative transfer
simulations favour a value for $\lambda_{\rm mfp}$ which is lower by a
around a factor of two at $z=5$ (\citealt{Gnedin04,GnedinFan06}).
However, halving our estimate for $\lambda_{\rm mfp}$ at $z=5$, we
still find that the observed galaxy and quasar population should
provide enough ionizing photons when assuming $M_{\rm AB}(1350)=-18$
with  a slightly larger escape fraction of $f_{\rm esc}\geq 0.3$.   At
$z=6$, our $\lambda_{\rm mfp}$ is comparable to the recent simulation
data, as well as the constraints on $\lambda_{\rm mfp}$ obtained by
\cite{Fan06}.  However, if $\lambda_{\rm mfp}$ is nevertheless
reduced by a factor of two, one instead requires $f_{\rm esc}\geq 0.4$
for consistency with the \cite{Bouwens06} luminosity function,
integrated to a limiting magnitude of $M_{\rm AB}(1350)=-18$.  These
escape fractions may be slightly high, but we think that given the
galaxy and quasar emissivities we derive are probably lower limits and
that the spectral shape of the sources is rather uncertain they are
not a cause for concern.    Note further that the re-emission of
ionizing photons by recombining ions, which has been neglected here,
may also raise the  ionization rate by as much as 50 per cent if the
spectrum of ionizing  photons is hard enough to significantly ionize
helium (\citealt{HaardtMadau96}).

\subsection{The clumping factor and the minimum ionizing emissivity
required to balance recombinations}

The emission rate of ionizing photons per unit comoving
volume, $\dot N_{\rm rec}$, required to balance recombinations in the
ionized part of the IGM is given by ({\it e.g.} MHR99)

\begin{equation} \dot N_{\rm rec} = \frac{ \langle n_{\rm H}
\rangle}{\langle t_{\rm rec} \rangle}  = 10^{50.0} C_{\rm HII} \left( \frac{1+z}{7}
\right)^{3} \rm~s^{-1}~Mpc^{-3}, \label{eq:Ncrit} \end{equation}

\noindent
where $\langle n_{\rm H} \rangle$ is the mean comoving hydrogen
density in the Universe and $\langle t_{\rm rec}\rangle $ is the
volume averaged recombination time for ionized hydrogen with an
effective \HII clumping factor $C_{\rm HII}=\langle n_{\rm HII}^{2}
\rangle/\langle n_{\rm HII} \rangle^{2}$.  The clumping factor
parameterises the inhomogeneity of the ionized hydrogen in the IGM and
therefore it is a redshift dependent quantity which increases with
decreasing redshift.  It does not include the contribution from
collapsed, virialised objects or clumps that are self-shielded from
the ionizing background.  Therefore, $C_{\rm HII}$ can be
significantly smaller than the total baryonic clumping factor.  MHR99
adopt $C_{\rm HII}=30$, based on the value at $z=5$ computed from the
cosmological radiative transfer simulation of \cite{GnedinOstriker97}.
\cite{Bunker04} also  adopt $C_{\rm HII}=30$ when computing the number
of ionizing photons needed to keep the IGM highly ionized at $z=6$, as
do a number of other authors.  However, a wide range of other values
for $C_{\rm HII}$ have been assumed in the literature (see
\citealt{Srbinovsky07} for details).  Note that we use the case-B
recombination coefficient evaluated at $10^{4}\rm~K$ in
equation~(\ref{eq:Ncrit}) ({\it e.g.}  MHR99).  However, if the case-A
recombination coefficient is the more appropriate choice, $\dot N_{\rm
rec}$ would be raised by a factor of $1.6$.  Alternatively, since the
recombination coefficient scales as $T^{-0.7}$, a gas temperature of
$2\times 10^{4}\rm~K$ would lower $\dot N_{\rm rec }$ by around the
same factor.  Lastly, we note equation~(\ref{eq:Ncrit}) can only
predict the number of ionizing photons required to balance
recombinations when the average recombination time is smaller than the
age of the Universe.

\begin{figure}
\begin{center}
 
  \includegraphics[width=0.45\textwidth]{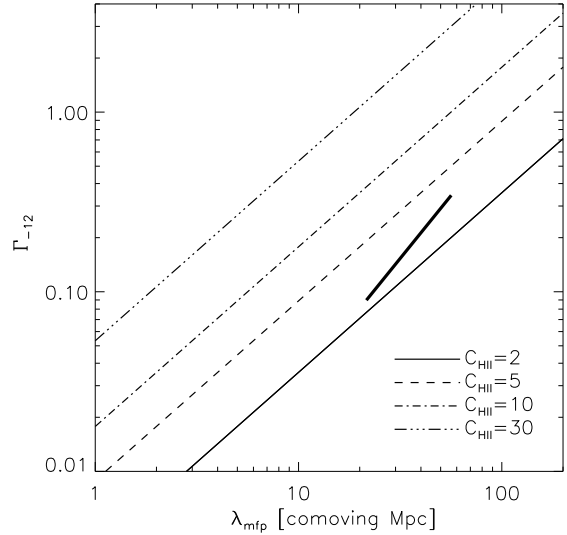}
  \caption{Comparison of our constraints on $\Gamma_{-12}$ and
  $\lambda_{\rm mfp}$ obtained from the IGM \Lya effective optical
  depth at $z=6$ (thick solid line) to the minimum hydrogen ionization
  rate required to ionize the IGM up to some density threshold
  corresponding to the clumping factor $C_{\rm HII}$ at $z=6$ (thin straight lines).}
\label{fig:critNion}
\end{center} 

\end{figure}

\begin{figure*}
  \centering 
  \begin{minipage}{180mm} 
    \begin{center}
           
      \psfig{figure=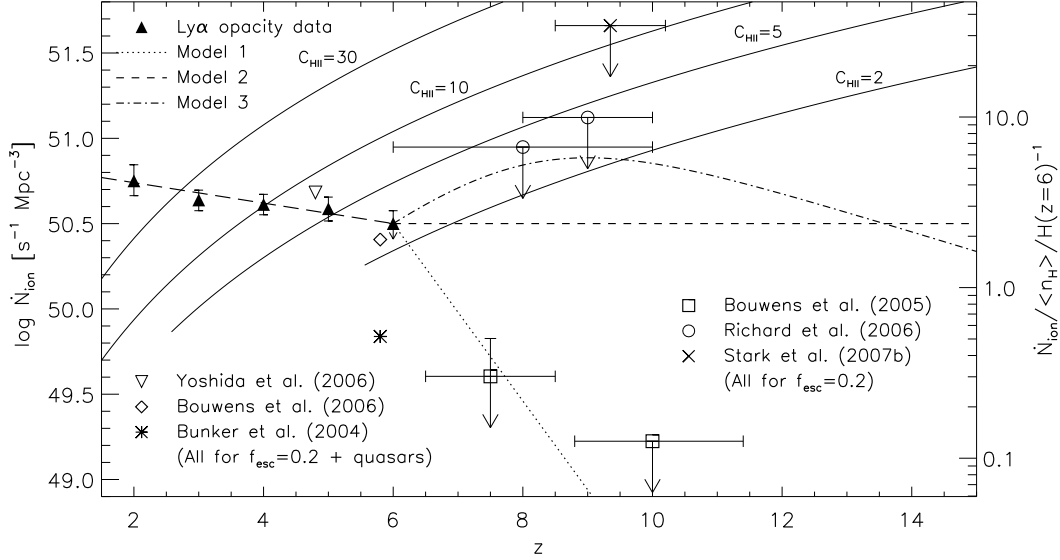,width=0.85\textwidth}
      \caption{Observational constraints on the emission rate of
      ionizing photons per comoving Mpc, $\dot N_{\rm ion}$, as a
      function of redshift.  All results are computed assuming
      $\alpha_{\rm s}=\alpha_{\rm b}=3$ and, in the case of the LBG
      and \Lya emitter emissivity estimates only, $f_{\rm
      esc}=0.2$. The scale on the right-hand vertical axis corresponds
      to the number of ionizing photons emitted per hydrogen atom over
      the Hubble time at $z=6$. The filled triangles give an estimate
      of $\dot N_{\rm ion}$ based on the constraints for
      $\Gamma_{-12}$ and $\lambda_{\rm mfp}$ obtained from the \Lya
      effective optical depth in this work and in B05.  The inverted
      triangle at $z=5$ and the diamond and star at $z=6$ correspond
      to estimates of $\dot N_{\rm ion}$ based on the Lyman limit
      emissivities of LBGs and quasars listed in Table 5.  The data
      have been slightly offset from their actual redshifts for
      clarity.  An escape fraction of $f_{\rm esc}=0.2$ has been
      assumed in this instance.  At $z>6$, the open squares and
      circles are derived from the upper limits on the comoving star
      formation rate per unit volume inferred by Bouwens et al. (2005)
      and Richard et al. (2006), respectively.  The cross is derived
      from the number density of \Lya emitters estimated by Stark et
      al. (2007).  Three simple models for the evolution of $\dot
      N_{\rm ion}$ are also shown as the dotted, short dashed and
      dot-dashed lines.  The solid lines correspond to the emission
      rate of ionizing photons per unit comoving volume, $\dot N_{\rm
      rec}$, needed to keep  the IGM ionized for various \HII clumping
      factors.  At $2 \leq z \leq 6$, we find $\dot N_{\rm ion}$ is
      characterised by a power law, $\dot N_{\rm ion}(z) =
      10^{50.5-0.06(z-6)}\rm~s^{-1}~Mpc^{-3}$, shown by the long
      dashed line.  Adopting a mean free path which is a factor of two
      smaller or setting $\alpha_{\rm s}=\alpha_{\rm b}=1$ will double
      the emissivity derived from the \Lya forest opacity.}
      \label{fig:nion} \end{center} \end{minipage}
\end{figure*}

The emission rate of ionizing photons can be converted to a
flux and hence a hydrogen ionization rate by assuming a photon mean free
path and a typical source spectral index, $\alpha_{\rm s}$.  The corresponding 
specific intensity at the Lyman limit is then ({\it e.g.} MHR99)

\begin{equation} J_{\rm L}^{\rm rec} = \frac{1}{4\pi}\lambda_{\rm
mfp} \dot N_{\rm
rec}h_{\rm p}\alpha_{\rm s}(1+z)^{2}, \end{equation}

\noindent
Assuming the spectral index of the ionizing background, $\alpha_{\rm
b}=\alpha_{\rm s}= 3$, the minimum ionization rate required to balance
recombinations in the IGM  with a clumping factor $C_{\rm HII}$ is

\[ \Gamma_{-12}^{\rm rec} \simeq 0.07~C_{\rm HII}
\left( \frac{\alpha_{\rm s}}{3}\right)\left(\frac{\alpha_{\rm b}+3}{6}\right)^{-1}
\left(\frac{\lambda_{\rm mfp}}{40 \rm ~ Mpc} \right) \]

\begin{equation} \hspace{5mm} \times \left( \frac{1+z}{7} \right)^{5}, \end{equation}

\noindent
although note that reprocessing of the intrinsic source spectrum by
the IGM can alter the spectral shape of ionizing background somewhat
(\citealt{HaardtMadau96}).  In Fig.~\ref{fig:critNion} we compare this
ionization rate required to balance recombinations for different \HII
clumping factors at $z=6$ (thin straight lines) to our determination
of $\Gamma_{\rm -12}$ and $\lambda_{\rm mfp}$ from the IGM \Lya
effective optical depth (thick solid line). Our result for the
photoionization rate is consistent with the value needed   to balance
hydrogen recombinations at $z=6$ if $C_{\rm HII} \leq 3$.  Adopting
$C_{\rm HII}=30$, a value of $\Gamma_{-12}$ substantially higher than
that inferred from the \Lya effective optical depth would be required
to keep the IGM highly ionized at $z=6$, as was already noted by
MHR99.  Even using a value for $\lambda_{\rm mfp}$ which is a factor
of two smaller, our constraints on $\Gamma_{-12}$ still require that
$C_{\rm HII}\leq7$ at $z=6$.  Comparing to some recent cosmological
radiative transfer simulations, \cite{Iliev06b} predict $C_{\rm
HII}<2$ at $z>11$, while \cite{Sokasian03} find $C_{\rm HII}\sim 4$ at
$z=3$.  However, simulations with higher spatial resolution may favour
larger values for $C_{\rm HII}$.   Nevertheless, we conclude adopting
$C_{\rm HII}=30$ overestimates the \HII clumping factor of the IGM at
$z=6$ by at least factor of $4$ and possibly by a factor of $10$.  We
therefore emphasise that care should be taken when drawing conclusions
from equation~(\ref{eq:Ncrit}) as $C_{\rm HII}$ is rather uncertain.

\section{The ionizing emissivity -- Evidence for a photon-starved and
extended period of reionization}

\subsection{The ionizing emissivity at $z = 2-6$}

In the last section we focussed on the hydrogen ionization rate and
the question of whether or not the expected ionizing flux from
observed high-redshift galaxies and quasars is sufficient to keep the
IGM highly ionized at $z=5-6$. We now turn again to the comoving
ionizing emissivity in order to discuss whether or not the inferred
emissivity, if extrapolated to higher redshift, is sufficient to
reionize hydrogen in the first place.

The filled triangles in Fig. 7 show the result of turning the
constraints on $\Gamma_{-12}$ from the \Lya opacity at $z=2-6$ derived here and
in B05 into an emission rate of ionizing photons per unit comoving
volume.  For this we use the relation

\[ \dot N_{\rm ion} \simeq
10^{51.2}\Gamma_{-12} \left(\frac{\alpha_{\rm
s}}{3} \right)^{-1} \left(\frac{\alpha_{\rm b}+3}{6}\right)
\left(\frac{\lambda_{\rm mfp}}{40\rm~Mpc}\right)^{-1} \]
\begin{equation} \hspace{5mm} \times \left(\frac{1+z}{7}\right)^{-2}
\rm~s^{-1}~Mpc^{-3}, \label{eq:gamma2N} \end{equation}

\noindent
which assumes the mean free path is much smaller than the horizon.
Note that this expression is independent of $f_{\rm esc}$, but does
depend inversely on the mean free path for ionizing photons.  We have
estimated $\lambda_{\rm mfp}$ at $z=2-6$ by using the simple model
discussed in Section 3 combined with the $\Gamma_{-12}$ constraints
from this work and B05.  The assumed values are summarised in Table 7
with a comparison to the horizon scale, and these are used to
calculate the filled triangles in Fig. 7.  The error estimates on
$\dot N_{\rm ion}$ computed using equation~(\ref{eq:gamma2N}) include
the error estimates for $\Gamma_{-12}$ and the corresponding error in
our estimate of the mean free path.  The comoving ionizing emissivity
inferred from the \Lya opacity is nearly constant over the redshift
range  $2 \leq z \leq 6$.  Assuming a spectral index $\alpha_{\rm
s}=\alpha_{\rm b}=3$ above the Lyman limit, it is well characterised
by a slowly rising power law, $\dot N_{\rm ion}(z) =
10^{50.5-0.06(z-6)}\rm~s^{-1}~Mpc^{-3}$,  shown by the long dashed
line.  A harder spectral index of  $\alpha_{\rm s}=\alpha_{\rm b}=1$
would double $\dot N_{\rm ion}$, while a smaller mean free path would
increase it linearly.

For comparison the inverted triangle, diamond and star at $z<6$  show
the estimates of  $\dot N_{\rm ion}$ based on the quasar and LBG
emissivities listed in Table 5 integrated to $M_{\rm AB}(1350)=-18$,
assuming $f_{\rm esc}=0.2$ and $\alpha_{\rm s}=3$.  As noted earlier,
$f_{\rm esc}\ga 0.2$ is required for consistency with our measurements
of $\Gamma_{-12}$ from the \Lya effective optical depth and estimates
of $\lambda_{\rm mfp}$.  The solid curves show  $\dot N_{\rm rec}$ as
given by equation~(\ref{eq:Ncrit}) for four different time independent
\HII clumping factors.  Note the curves are cut off at the redshift
where the recombination time exceeds the age of the Universe.   We
have not attempted to estimate errors for these estimates of the
ionizing emissivity, but note again that the assumed LBG spectral
shape is rather uncertain and depends strongly on the IMF, metallicity
and star formation history. Together with the uncertainty on the
escape fraction this leads to an error which may well approach an
order of magnitude.  In contrast, the estimates from the \Lya opacity
are significantly more robust. The biggest uncertainties associated
with these data are the Lyman limit spectral index and the mean free
path of ionizing photons, which introduce an uncertainty of around a
factor of 2-3 (see \citealt{Meiksin05} for a detailed discussion of
this point).

Finally,  it is illustrative to express   $\dot N_{\rm ion}$ in terms
of the number of ionizing photons emitted per hydrogen atom per time
interval (\citealt{MiraldaEscude03}).  This quantity is shown on the
right vertical axis in Fig.7 for a time period corresponding to the
Hubble time at $z=6$.   For $\alpha_{\rm s}=\alpha_{\rm b}=3$ the
comoving ionizing emissivity  inferred from the \Lya opacity
corresponds to the emission of $\sim 1.5$ photons per hydrogen atom in
a period that corresponds to the age of the Universe at $z=6$. For
$\alpha_{\rm s}=\alpha_{\rm b}=1$ this number would be twice that.
Reionization must therefore have occurred  in a  {\it photon-starved}
manner unless the ionizing emissivity was much higher at $z>6$.

\subsection{Three simple models for the ionizing emissivity of galaxies at $z>6$}

We now proceed to discuss plausible extrapolations of the ionizing
emissivity towards higher redshift. For this we firstly consider the
recent tentative upper limits on the UV emissivity from searches for
high-redshift galaxies at $z>6$.  Estimates based on the Hubble Ultra
Deep Field yield $\epsilon(1500)<4.0\times
10^{25}\rm~erg~s^{-1}~Hz^{-1}~Mpc^{-3}$ at $z\sim 7.5$ and
$\epsilon(1500)<1.0\times 10^{25}\rm~erg~s^{-1}~Hz^{-1}~Mpc^{-3} $ at
$z\sim 10$ (\citealt{Bouwens05}).  Similar results have also been recently
reported by \cite{Mannucci07} at $z\sim 7$. Alternative constraints from near
infra-red observations around lensing clusters give higher values,
$\epsilon(1500)<5.3\times 10^{26}\rm~erg~s^{-1}~Hz^{-1}~Mpc^{-3}$ at
$z\sim 8$ and $\epsilon(1500)<7.9\times 10^{26}
\rm~erg~s^{-1}~Hz^{-1}~Mpc^{-3} $ at $z\sim 9$ (\citealt{Richard06}).
These data omit a correction for dust extinction and are computed by
integrating the LBG luminosity function to a lower luminosity limit of
$0.3L_{*}(z=3)$.  In addition, recent searches for faint
gravitationally lensed \Lya emitters at $8.5<z<10.2$ behind foreground
galaxy clusters, presented by
\cite{Stark07b}, have given an upper
limit\footnote{Estimated from the upper limit on the number of sources
with a \Lya luminosity brighter than $L$, presented in fig. 11 of
Stark et al. (2007b).} of around $\epsilon_{\rm Ly\alpha} \la 2 \times
10^{41} \rm~erg~s^{-1}~Mpc^{-3}$, assuming all six of their
candidates are real.

The upper limits for $\epsilon(1500)$ can be related to the
emission rate of ionizing photons per unit comoving volume, $\dot
N_{\rm ion}$, by

\begin{equation} {\dot N}_{\rm ion} \simeq
\frac{\epsilon_{\rm L}f_{\rm esc}}{h_{\rm p}\alpha_{\rm s}} =
10^{49.7}\epsilon_{25}^{\rm g} 
\left(\frac{\alpha_{\rm s}}{3}\right)^{-1} \left(\frac{f_{\rm esc}}{0.1}\right)
\rm~s^{-1}~Mpc^{-3}, \label{eq:rhosfr} \end{equation}

\noindent
where we again assume $\epsilon_{\rm L}=\epsilon(1500)/6$ and a source
spectral index of $\alpha_{\rm s}=3$.  To convert the \Lya emissivity
to $\dot N_{\rm ion}$ we use,
 
\[ {\dot N}_{\rm ion} \simeq \frac{3}{2}\frac{f_{\rm
esc}^{\rm Ly\alpha}}{1-f_{\rm esc}} \frac{\epsilon_{\rm Ly\alpha}}{h_{\rm p}\nu_{\rm
Ly\alpha}} \]
\begin{equation} \hspace{6mm} =  10^{51.3} \epsilon_{41}
\left(\frac{f_{\rm esc}^{\rm Ly\alpha}}{0.2} \right) \left( \frac{1-f_{\rm esc}}{0.9}\right)^{-1} \rm~s^{-1}~Mpc^{-3}, \end{equation}

\noindent
where $\epsilon_{41}=\epsilon_{\rm
Ly\alpha}/10^{41}\rm~erg~s^{-1}~Mpc^{-3}$, $f_{\rm esc}^{\rm
Ly\alpha}$ is the escape fraction of \Lya photons and we assume case B
recombination, corresponding to the production of two \Lya photons for
every three ionizing photons (\citealt{Osterbrock89}).  We adopt
$f_{\rm esc}^{\rm Ly\alpha}=0.2$ based on the lower limit inferred
from the observations of \cite{Gawiser06} at $z\simeq 3$.

\begin{table} 
\centering 

\caption{Summary of $\Gamma_{-12}$ inferred from the \Lya forest
effective optical depth at $2\leq z \leq 6$.  The data at $z=2-4$ are
taken from B05, while the $z=5-6$ values are based on this work.  The
corresponding estimates for the mean free path in units of comoving
Mpc are obtained using the model discussed in Section 3.  For
comparison, we also list the horizon scale in comoving Mpc, $r_{\rm hor}$,
computed to three significant figures for our fiducial cosmology.}

  \begin{tabular}{c|c|c|c}
    \hline
    z     & $\Gamma_{-12}$       & $\lambda_{\rm mfp}$ & $r_{\rm
    hor}$ \\           
  \hline
  2     & $1.29^{+0.80}_{-0.46}$  & $770^{+233}_{-166}$ & $9360$ \\
 & &\\
  3     & $0.86^{+0.34}_{-0.26}$  & $374^{+82}_{-73}$ & $8140$  \\
	& &\\
  4     & $0.97^{+0.48}_{-0.33}$  & $286^{+87}_{-69}$ & $7290$ \\
& &\\
  5     & $0.52^{+0.35}_{-0.21}$  & $114^{+49}_{-34}$ & $6660$ \\
& &\\
  6     & $<0.19$  & $<37$ & $6170$ \\
   \hline
\end{tabular}
\end{table}

Three simple models for the evolution of $\dot N_{\rm ion}$ at $z>6$
are displayed along with these data points in Fig.7.  The models are
chosen simply to bracket a range of plausible evolutionary histories
of $\dot N_{\rm ion}$ at $z>6$, which are anchored at our  measurement
at $z=6$.  Model 1, shown by the dotted line, is a declining
power law consistent with the \cite{Bouwens05} data.   Model 2,
corresponding to the short dashed line, assumes $\dot N_{\rm ion}$
remains constant at $z>6$.  Lastly, the dot-dashed line corresponds to
model 3, a double exponential which peaks at $z=9$.  Note that recent
estimates of the mass assembled in star forming galaxies at $z\simeq
6$ indicate such an increase may be plausible
(\citealt{Mobasher05,Eyles06,Stark07a,Yan06}).  Alternatively a
distinct population of high redshift ionizing sources such as
population-III stars or mini-quasars ({\it e.g.}
\citealt{Venkatesan03,Sokasian04,Madau04,Ricotti05,ChoudhuryFerrara05,Meiksin05})
may be responsible for an increased ionizing emissivity at high
redshift.   For all three models we assume $\dot N_{\rm ion}=0$ at
$z>15$ and $\dot N_{\rm ion}(z) =
10^{50.5-0.06(z-6)}\rm~s^{-1}~Mpc^{-3}$ at $z<6$.

\begin{figure}
\begin{center} 
  \includegraphics[width=0.45\textwidth]{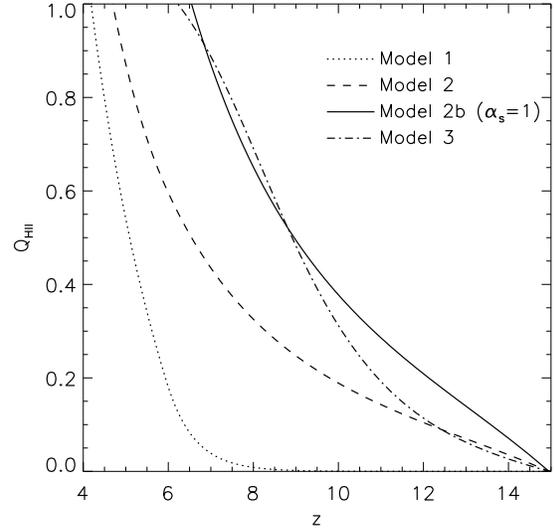}
\caption{The \HII filling factor as a function of redshift computed
using the three models for the redshift evolution of $\dot N_{\rm
ion}$ shown in Fig. 7.  A time independent \HII clumping factor of
$C_{\rm HII}=2$ and a source spectral index of $\alpha_{\rm s}=3$
has been assumed in this instance.  The solid line shows the filling
factor expected for model 2 if $\dot N_{\rm ion}$ is raised by a
factor of two, corresponding to a source spectral index of
$\alpha_{\rm s}=1$.}
\label{fig:fill}
\end{center} 
\end{figure}

We can now obtain the filling factor of \HII in the IGM for the three
models for $\dot N_{\rm ion}$ by solving ({\it e.g.}
MHR99,\citealt{WyitheLoeb03,ChoudhuryFerrara05})

\begin{equation} \frac{dQ_{\rm HII}}{dt} = \frac{{\dot N}_{\rm
ion}}{\langle n_{\rm H} \rangle} - Q_{\rm HII}C_{\rm
HII} \frac{\langle n_{\rm H} \rangle}{a^{3}}\alpha_{\rm B}(T), \end{equation}

\noindent
where $\alpha_{\rm B}(T)$ is the case-B recombination coefficient,
which we evaluate at $T=10^{4}\rm~K$, $\langle n_{\rm H} \rangle$ is
the mean comoving hydrogen density and $a=(1+z)^{-1}$ is the
cosmological expansion factor.  We take $Q_{\rm HII}=1$ as a proxy for
the end point of hydrogen reionization.  This occurs at progressively earlier
redshifts for either a larger value of $\dot N_{\rm ion}$ or a smaller
value of $C_{\rm HII}$.  The \HII filling factor is shown as a
function of redshift for our three simple models for the evolution of 
$\dot N_{\rm ion}$ in Fig.~\ref{fig:fill}.  Assuming a time independent \HII clumping
factor of $C_{\rm HII}=2$, $Q_{\rm HII}=1$ is reached at $z=4.16$ for
model 1, $z=4.69$ for model 2 and $z=6.22$ for model 3.  Hydrogen
reionization is only complete by $z=6$ in model 3, and is the only
model consistent with the observed flux distribution in the spectra of
high redshift quasars (\citealt{Fan06,Becker07}).

However, before proceeding we should discuss  the two main
uncertainties involved in calculating $Q_{\rm HII}$ from our three
models; the source spectral index $\alpha_{\rm s}$ and the clumping
factor $C_{\rm HII}$.  The filling factors so far were obtained
assuming a source spectral index of $\alpha_{\rm s}=3$.  As discussed
in some detail by \cite{Meiksin05}, a harder spectral index will
increase $\dot N_{\rm ion}$. As an example, adopting $\alpha_{\rm
s}=\alpha_{\rm b}=1$ doubles the $\dot N_{\rm ion}$ we infer at our
anchor point of $\Gamma_{-12}$ at $z=6$.  The solid line labelled
model 2b in Fig.~\ref{fig:fill} shows $Q_{\rm HII}$ computed for a
constant $\dot N_{\rm ion}$ twice that in model 2.  The resulting
$Q_{\rm HII}$ evolution is very similar to model 3 and is now
consistent with hydrogen reionization being complete by $z=6$. On the
other hand, adopting a softer spectral index of $\alpha_{\rm s}=5$
({\it e.g.}  \citealt{BarkanaLoeb01}) would push reionization to even
later times, requiring an even larger increase in $\dot N_{\rm ion}$
to achieve $Q_{\rm HII}=1$ by $z=6$.  Our adopted value for the
clumping factor at is less of a concern.  Even if we assume a uniform
IGM ($C_{\rm HII}=1$), model 2 still only predicts $Q_{\rm HII}=1$ by
$z=5.26$.  In addition, the clumping factor will become larger towards
lower redshifts.

Finally, we note that a particularly hard spectral index has
implications for the \HeII reionization history.  The much larger
ionization threshold of singly ionized helium relative to neutral
hydrogen ensures that \HeII reionization is postponed until sources
with sufficiently hard spectra become numerous.  Quasars are likely to
be the primary source of these energetic photons, and there is some
evidence that the tail-end of \HeII reionization does roughly coincide
with the peak in the quasar number density around $z\sim 2- 3$
(\citealt{Shull04,Bolton06}).  However, a spectral index of
$\alpha_{\rm s}=1$ at $z\geq6$ would result in an earlier \HeII
reionization epoch, at odds with the current observational evidence.
It may therefore be possible to rule this scenario out. 

\subsection{The IGM neutral hydrogen fraction and electron scattering
optical depth}

\begin{figure*}
  \centering 
  \begin{minipage}{180mm} 
    \begin{center}
           
      \psfig{figure=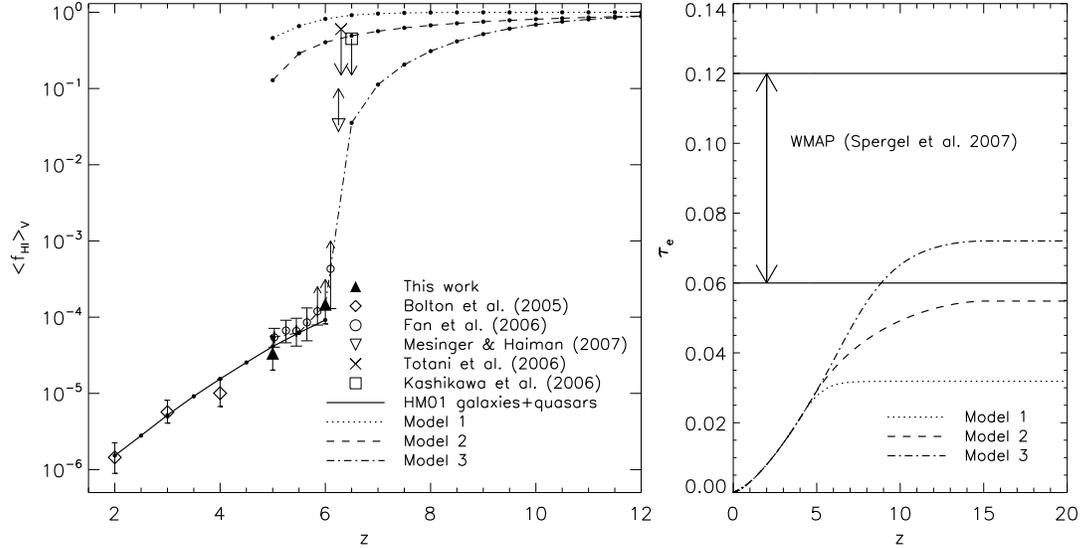,width=0.85\textwidth}
      \caption{ {\it Left:}  Evolution of the volume weighted neutral
      hydrogen fraction in the IGM as a function of redshift.  The
      filled triangles and open diamonds at $z \leq 6$ correspond to
      the $\langle f_{\rm HI} \rangle_{\rm V}$ computed using the
      constraints on $\Gamma_{-12}$ obtained in this work and in B05,
      respectively.  The solid line gives the $\langle f_{\rm HI}
      \rangle_{\rm V}$ calculated using the UV background model of
      HM01 for galaxies and quasars.  Other observational constraints
      at $z>6$ come from Mesinger \& Haiman (2007), Totani et
      al. (2006) and Kashikawa et al. (2006), shown by the inverted
      triangle, cross and square respectively.  The expected evolution
      in $\langle f_{\rm HI} \rangle_{\rm V}$ at $ z>5$ for the three
      models for $\dot N_{\rm ion}$ shown in Fig.~\ref{fig:nion} are
      shown as the dotted, dashed and dot-dashed lines.  {\it Right:}
      The cumulative electron scattering optical depth as a function
      of redshift computed from the three models for $\dot N_{\rm
      ion}$ shown in Fig.~\ref{fig:nion}. } \label{fig:fH1} \end{center}
      \end{minipage}
\end{figure*}

The \HII filling factors calculated in the last section may be used 
to estimate the volume weighted
neutral fraction in the IGM for the proposed $\dot N_{\rm ion}$ models by solving

\begin{equation} \langle f_{\rm HI} \rangle_{\rm V} = 1 + Q_{\rm
HII}\left[\frac{\int_{0}^{\Delta_{\rm IGM}} d\Delta~f_{\rm
HI}(\Delta) P_{\rm V}(\Delta)}{\int_{0}^{\Delta_{\rm
IGM}}d\Delta~P_{\rm V}(\Delta)} -1\right], \label{eq:fH1} \end{equation}

\noindent
when $Q_{\rm HII}< 1$ and

\begin{equation} \langle f_{\rm HI} \rangle_{\rm V} =
\frac{\int_{0}^{\Delta_{\rm IGM}} d\Delta~f_{\rm HI}(\Delta) P_{\rm V}(\Delta)}{\int_{0}^{\Delta_{\rm IGM}}d\Delta~P_{\rm V}(\Delta)}, \label{eq:fH1b} \end{equation}

\noindent
when $Q_{\rm HII}\geq 1$.  Here $f_{\rm HI}(\Delta)$ is the residual
neutral hydrogen fraction in an ionized region with overdensity
$\Delta$ and $\Delta_{\rm IGM}$ is the upper limit for the gas
overdensity in the IGM, which we take to be $\Delta_{\rm IGM}=150$
(\citealt{Fan06}).   We compute $f_{\rm HI}(\Delta)$ at a particular
overdensity assuming ionization equilibrium, and then integrate over
the volume weighted probability distribution of the gas density,
$P_{\rm V}(\Delta)$.  Since there are no measurements of
$\Gamma_{-12}$ at $z>6$ we have  no independent constraints on the
mean free path.  We therefore adopt an upper limit of $\lambda_{\rm
mfp} = 20$ comoving Mpc at $z>6$. In practice, since $Q_{\rm HII}=1$
occurs at $z<6.25$ for all three models, changing the mean free path
(or indeed $P_{\rm V}(\Delta)$, which is also rather uncertain) does
not substantially change the $\langle f_{\rm HI} \rangle_{\rm V}$ we
derive.  As already discussed, the main uncertainties in the
calculation at $z>6$ are the values adopted for $\alpha_{\rm s}$ and
$C_{\rm HII}$.

The evolution of $\langle f_{\rm HI} \rangle_{\rm V}$ as a function of
redshift is shown in the left panel of Fig.~\ref{fig:fH1}.  The
constraints on $\langle f_{\rm HI} \rangle_{\rm V}$ at $2\leq z\leq
6$, shown by the filled triangles and open diamonds, are computed
using the ionization rates inferred from the IGM \Lya effective
optical depth in this
work and in B05, respectively.  When doing this we assume $Q_{\rm
HII}=1$ in equation~(\ref{eq:fH1}); there is strong evidence the
hydrogen in the IGM is already highly ionized at $z\leq6$
(\citealt{Fan06,Becker07}).  The values we infer for the volume
weighted neutral fraction are in good agreement with the recent data
from \cite{Fan06} at $z \simeq 6$, shown by the open circles.  The
solid line corresponds to $\langle f_{\rm HI} \rangle_{\rm V}$
computed at intervals of $\Delta z = 0.5$ using the updated ionizing
background model of HM01 for galaxy and quasar emission.  There is
also very good agreement between this model and the data based on the
IGM \Lya effective optical depth.

Observational constraints on $\langle f_{\rm HI} \rangle_{\rm V}$ at
$z>6$ using alternative observational techniques are also shown.  The
cross in the left panel of Fig.~\ref{fig:fH1} corresponds to the upper
limit on $\langle f_{\rm HI} \rangle_{\rm V}$ at $z=6.3$ obtained from
the spectrum of GRB 050904 by \cite{Totani05}.  The data point at
$z=6.25$, represented by the inverted triangle, shows the recent lower
limit on $\langle f_{\rm HI} \rangle_{\rm V}$ measured by
\cite{MesingerHaiman07} using the spectroscopically observed sizes of
near-zones around quasars combined with modelling of the Gunn-Peterson
trough damping wing.  Note, however, that the estimates of the neutral
fraction based on the observed near-zone sizes alone are ambiguous
(\citealt{BoltonHaehnelt07,Maselli07}); the near-zone data is also
consistent with an IGM which is  highly ionized at $z\sim 6$.  The
open  square corresponds to the upper limit on $\langle f_{\rm HI}
\rangle_{\rm V}$ obtained by \cite{Kashikawa06} from the observed
evolution in the \Lya emitter luminosity function between $z=5.7$ and
$z=6.5$.   However, this estimate is also subject to considerable
uncertainties  (\citealt{Santos04,Dijkstra07}).

The estimates for $\langle f_{\rm HI} \rangle_{\rm V}$  computed with
equations~(\ref{eq:fH1}) and (\ref{eq:fH1b}) for the three $\dot
N_{\rm ion}$ models at $z>5$  are shown as the dotted, dashed and
dot-dashed lines.  The \HII filling factor, $Q_{\rm HII}$, is computed
assuming $C_{\rm HII}=2$.  Model 1 is inconsistent with all the
observational constraints on $\langle f_{\rm HI} \rangle_{\rm V}$ at
$z>5$.  Similarly, although model 2 predicts a somewhat smaller
$\langle f_{\rm HI} \rangle_{\rm V}$ at $z>5$, hydrogen reionization
is still not complete by $z=6$, in disagreement with the neutral
fraction inferred from the IGM \Lya effective optical depth.   Hence
the values of $\langle f_{\rm HI} \rangle_{\rm V}$  predicted by both
model 1 and 2 lie well above the measurements from the \Lya data at
$z=5-6$.  Only model 3 (and also 2b, not shown) is in agreement with
the \Lya data.

We may also briefly discuss to what extent the three models  are
consistent with current constraints on the integrated reionization
history from the cosmic microwave background.  The recently
re-evaluated value for the electron scattering optical depth is
$\tau_{\rm e}=0.09\pm0.03$ (\citealt{Spergel06}).  The electron
scattering optical depth  may be computed as ({\it e.g.}
\citealt{WyitheLoeb03,ChoudhuryFerrara05})

\begin{equation} \tau_{\rm e} = c\sigma_{\rm T}\langle n_{\rm H}
\rangle \int_{0}^{z^{\prime}}dz~Q_{\rm
HII}(z)(1+z)^{3}\left|\frac{dt}{dz}\right|, \end{equation}

\noindent
where $\sigma_{\rm T}=6.65 \times 10^{-25}\rm~cm^{2}$ is the Thomson
cross-section.  The cumulative contribution to $\tau_{\rm e}$ as a
function of redshift is shown in the right hand panel of
Fig.~\ref{fig:fH1} for our three models.  As might be expected,
only model 3 is consistent with the current constraints on $\tau_{\rm
e}$.  Note again that we have arbitrarily assumed that the ionizing
emissivity is zero at $z>15$.  Obviously for model 2 and 3
the value of $\tau_{\rm e}$ would be larger for an earlier start
of reionization.  Free electrons left after recombination may also
increase $\tau_{\rm e}$ somewhat (see \citealt{Shull07} for a recent
detailed discussion of these issues).

\subsection{A photon-starved extended epoch of reionization:
Implications  for 21cm experiments and the photo-evaporation of mini-haloes}

Even for a  hard spectral index for ionizing photons, $\alpha_{\rm
s}\sim 1$, the comoving ionizing emissivity at $z=6$ inferred from the
\Lya data  is so low that the epoch of reionization has to extend over
a wide range  in redshift.  This is obviously excellent news for
upcoming 21cm experiments as there should be plenty of structure
observable in 21cm emission over the full accessible frequency range of
planned experiments.   It appears likely  that  a significant fraction
of the IGM is still neutral at $z=8$, which would bode particularly
well for observations of \HII regions around early ionizing  sources
by upcoming 21cm experiments such as  LOFAR  ({\it e.g}
\citealt{WyitheLoebBarnes05,Zaroubi05, RhookHaehnelt06}).  The only
downside regarding 21cm experiments is that there is little hope for
those experiments aimed at  a detection of a global,  sharp
reionization signal (\citealt{Shaver99}).

An extended photon-starved epoch of reionization also has important
implications for the photo-evaporation of mini-haloes, {\it i.e.} dark
matter haloes with virial temperatures below $10^{4}\rm~K$. The virial
temperature of these haloes is too low for collisional cooling by
atomic hydrogen to be efficient. Prior to reionization they are thus
expected to be  filled with neutral hydrogen.  However, after
reionization commences they will be photo-evaporated as
photoionization  raises the gas temperature above the virial
temperature of the halo.  The number of photons per hydrogen atom in
the halo needed for  photo-evaporation, and hence the duration of the
photo-evaporation process itself, will therefore depend on the number
of recombinations occurring.  \cite{Iliev05} studied this in
detail for a large suite of radiative transfer simulations. For haloes
at the upper end of virial temperatures with total masses of about
$10^{7}M_{\odot}$ and photo-evaporating ionizing fluxes that
corresponds to our measured photoionization rate at $z=6$, the time
needed for  photo-evaporation  approaches a Gyr.  Unless the ionizing
emissivity rises dramatically at $z>6$ the photoionization rate
towards higher redshift will be even lower. It should therefore take
until about $z=3-4$ to fully photo-evaporate the most massive
mini-haloes. This would support the suggestion of \cite{AbelMo98}
that a significant fraction of observed Lyman limit systems, which
increase in number with increasing redshift, are due to optically
thick neutral hydrogen in mini-haloes.

\section{Summary and conclusions}

We have obtained new measurements of the metagalactic hydrogen
ionization rate at $z=5$ and $6$ using a large suite of hydrodynamical
simulations combined with recent measurements of the IGM \Lya
effective optical depth (\citealt{Songaila04,Fan06}).  We carefully
take into account the various systematic errors associated with
determining $\Gamma_{-12}$ at these redshifts.  We find $\log
\Gamma_{\rm HI} = -12.28^{+0.22}_{-0.23}$ at $z=5$ and $\log
\Gamma_{\rm HI} < -12.72$ at $z=6$, where the largest contributions to
the uncertainties come from the poorly defined thermal state of the
IGM and measurements of the \Lya effective optical depth.  We also
investigate the impact of ionizing background fluctuations on the
inferred values of $\Gamma_{-12}$, and find that these add an
additional uncertainty of about $10$ per cent.

Using a physically motivated model for the ionizing photon mean free
path, we compare our constraints on $\Gamma_{-12}$ to the expected
ionization rate calculated from the observed galaxy and quasar
population at $z=5$ and $6$.  We find that, even for conservative
estimates regarding the spectral shape of the UV emission of these
sources, the combined ionizing emission from galaxies and quasars is
capable of maintaining the IGM in its highly ionized state if $f_{\rm
esc}\ga 0.2$.  Ionizing emission from star forming galaxies is likely
to dominate the total ionizing photon budget at these redshifts.  The
clumping factor of ionized hydrogen should also be  substantially
less than the often used value of $C_{\rm HII}=30$ at $z=5-6$.
Our measurements of  $\Gamma_{-12}$ and estimates of $\lambda_{\rm
mfp}$ suggest  $C_{\rm HII}\la 3$ at $z=6$, although the clumping
factor will increase towards lower redshift.  

Using  our estimates of the  ionizing photon mean free path we have
turned our measurements of  $\Gamma_{-12}$ from the \Lya effective
optical depth into a measurement of the emission rate of ionizing
photons per unit comoving volume at $2\leq z \leq 6$.   We find that
$\dot N_{\rm ion}(z) = 10^{50.5-0.06(z-6)}\rm~s^{-1}~Mpc^{-3}$
assuming $\alpha_{\rm s}=\alpha_{\rm b}=3$ above the Lyman limit
frequency. A harder spectral index of $\alpha_{\rm s}=\alpha_{\rm
b}=1$ would double $\dot N_{\rm ion}$.  For $\alpha_{\rm
s}=\alpha_{\rm b}= 3$ and $1$  the value of $\dot N_{\rm ion}$ at
$z=6$ corresponds to around $1.5$ and $3$ photons emitted per hydrogen
atom over a time interval corresponding to the age of the Universe at
$z=6$, respectively.  Reionization  must therefore have occurred in a
photon-starved manner unless the ionizing emissivity during
reionization is substantially larger than at $z=2-6$.

We have discussed three simple extrapolations of $\dot N_{\rm ion}$ at
$z>6$.  An ionizing emissivity which decreases rapidly towards higher
redshift,  consistent with recent upper limits on the UV emissivity
from LBGs at $z>6$ (\citealt{Bouwens05}), can be ruled out based on
the ionization state of the IGM at $z<6$.  Consistency with the
completion of reionization before $z=6$ as inferred from  the $\langle
f_{\rm HI} \rangle_{\rm V}$ in the IGM, as well as the recently
re-determined electron scattering  optical depth, is achievable if the
ionizing emissivity at $z>6$ increases above its inferred value at
$z=6$.  Alternatively, consistency may also be achieved if the
$1500\rm\AA$ luminosity density and the star formation rate remain
constant while $f_{\rm esc}$ increases at $z>6$, or the  spectrum of
emitted ionizing photons becomes increasingly harder at high redshift.
Recent studies of LBGs at $z= 5$ suggest that typical UV selected
star-forming galaxies become younger with increasing redshift
(\citealt{Verma07}), which may be a hint in this direction.
Reionization by an early population of sources such as mini-quasars or
population-III stars with  particularly hard spectra ({\it e.g.}
\citealt{Venkatesan03,Sokasian04,Madau04,Ricotti05}) would be an
extreme form of such a scenario.

Assuming $\epsilon(1500)=6\epsilon_{\rm L}$ ({\it e.g.} MHR99), a
constant ionizing emissivity of around  $\epsilon(1500)\sim  3.8
\times 10^{26} (\alpha_{\rm s}/3)(f_{\rm
esc}/0.2)^{-1}\rm~erg~s^{-1}~Hz^{-1}~Mpc^{-3}$ is required for
reionization before or at $z=6$.  This is at the upper end of some of
the recent first attempts to constrain   the UV luminosity density  at
$z>6$ (\citealt{Bouwens05,Richard06,Mannucci07,Stark07b}).  There is thus reason
for optimism regarding future  searches for the UV emission from
high-redshift galaxies. Note, however, that the required $1500\rm\AA$
luminosity density could be smaller if the UV spectrum of the galaxies
is harder or the escape fraction is larger than we have assumed.

The photon-starved nature of reionization suggested by the \Lya forest
data at redshift $z=5$ and $z=6$ means that the epoch of reionization
has to extend over a wide redshift range and is unlikely to have been
completed much before $z=6$. In turn this implies a rather early start
of reionization as suggested independently by the  CMB data.  There
should thus be plenty of structure due to the epoch of reionization
observable in 21cm emission over the full accessible frequency range
of planned 21cm experiments. The photon-starved nature of reionization
also means that  the photo-evaporation of the neutral hydrogen in
mini-haloes with virial  temperatures below $10^4 {\rm K}$ should
extend to redshifts well below the tail-end of the hydrogen
reionization epoch.  The alternative is that the IGM was rapidly
reionized at very high redshift by an as yet unidentified population
of sources which have disappeared by $z\sim6$.  

\section*{Acknowledgements}

We are grateful to Francesco Haardt for making his updated UV
background model available to us.  We also thank Benedetta Ciardi and
Tirth Choudhury for comments on the draft manuscript, and the
anonymous referee for a detailed report which helped to improve
this paper.  This research was conducted in cooperation with SGI/Intel
utilising the Altix 3700 supercomputer COSMOS at the Department of
Applied Mathematics and Theoretical Physics in Cambridge.  COSMOS is a
UK-CCC facility which is supported by HEFCE and PPARC.

\end{document}